 \documentclass[sigconf]{acmart}

\AtBeginDocument{%
  \providecommand\BibTeX{{%
    \normalfont B\kern-0.5em{\scshape i\kern-0.25em b}\kern-0.8em\TeX}}}

\copyrightyear{2025}
\acmYear{2025}
\setcopyright{cc}
\setcctype{by}
\acmConference[CHI '25]{CHI Conference on Human Factors in Computing Systems}{April 26-May 1, 2025}{Yokohama, Japan}
\acmBooktitle{CHI Conference on Human Factors in Computing Systems (CHI '25), April 26-May 1, 2025, Yokohama, Japan}\acmDOI{10.1145/3706598.3713243}
\acmISBN{979-8-4007-1394-1/25/04}

\begin{document}

\title[Practitioner Challenges and Experiences of Visualizing Race and Gender]{The Social Construction of Visualizations:\\ Practitioner Challenges and Experiences of Visualizing \\ Race and Gender Demographic Data}

\author{Priya Dhawka}
\email{pdhawka@uw.edu}
\orcid{0000-0002-7334-4950}

\affiliation{%
  \institution{University of Washington}
  \city{Seattle}
  \state{Washington}
  \country{USA}
}

\author{Sayamindu Dasgupta}
\email{sdg1@uw.edu}
\orcid{0000-0001-6083-2114}
\affiliation{%
  \institution{University of Washington}
  \city{Seattle}
  \state{Washington}
  \country{USA}}

 \renewcommand{\shortauthors}{Dhawka and Dasgupta}

\begin{abstract}
Data visualizations are increasingly seen as socially constructed, with several recent studies positing that perceptions and interpretations of visualization artifacts are shaped through complex sets of interactions between members of a community. However, most of these works have focused on audiences and researchers, and little is known about if and how practitioners account for the socially constructed framing of data visualization. In this paper, we study and analyze how visualization practitioners understand the influence of their beliefs, values, and biases in their design processes and the challenges they experience. In 17 semi-structured interviews with designers working with race and gender demographic data, we find that a complex mix of factors interact to inform how practitioners approach their design process---including their personal experiences, values, and their understandings of power, neutrality, and politics. Based on our findings, we suggest a series of implications for research and practice in this space.
\end{abstract}






\maketitle

\section{Introduction}
In March 2024, the United States (US) government approved the addition of ``Middle Eastern or North African'' (MENA) as a new racial or ethnic category on future government data collection forms~\cite{MENA2024}. This change in racial classification came after years of campaigning on behalf of activists and individuals who previously were classified as ``White'' according to US census racial categories~\cite{MENA-NYT}. Although the US census will not reflect this change until 2030, once in effect, this new demographic category change can greatly impact public policy, voting districts, and access to resources more generally in the US~\cite{MENA-NYT}. Due to its socially constructed nature, equitably visualizing demographic data has recently been a research challenge of interest, especially when representing marginalized populations through anthropographics (human-shaped visualizations) of demographic data~\cite{Boy, Morais2020,dhawka2023we}. This interest reflects increasing acknowledgment within the visualization research community that data and visualizations are the product of human choices and are thus not neutral.  

A growing body of work has examined the various factors influencing the creation of data visualizations, including designer choices, tools, and creative practices~\cite{parsons2021fixation, parsons2021understanding, bako2022understanding}. There has been an influx of work that explores how feminist, critical, and alternative epistemologies ~\cite{dork,datafem, Correll, akbaba2023troubling, Schwan2022, akbaba2024entanglements, Berret-Munzner2024} can supplement mainstream positivist approaches in visualization research. Much of this work has largely focused on researchers and consumers of visualizations or the visualization artifacts as the units of analysis and have mostly been data-agnostic, that is, have not considered the unique questions that accompany specific kinds of data. To address this gap, in this paper, we focus on visualization designers and the issues that they grapple with as they produce visualizations of socially constructed demographic data such as race and gender.

Demographic data, especially in the US context, generally refers to information about personal, individual characteristics such as age, gender, race, and socioeconomic information, often collected in relation to education, employment, income levels, birth, and death rates. Demographic data collection frequently involves grouping individuals into distinct population groups using physical or arbitrary characteristics, many of which can be historically linked to experiences of marginalization or discrimination~\cite{scheuerman2021datasets, baah2019marginalization, pine2015politics}. In this work, we focus on race and gender demographic data which the United States Equal Employment Opportunity Commission (EEOC) defines as protected characteristics~\cite{USEEOC} and we refer to this data as protected demographic data throughout this paper. In the US, it is illegal to discriminate based on protected categories such as race, color, religion, sex, national origin, disability, and genetic information~\cite{ProtectedDem}. Visualizing demographic data is challenging as demographic categories are often fluid, constructed from arbitrary measures that are influenced by sociopolitical factors, can change over time, and can be closely tied to political outcomes for marginalized groups~\cite{Andrus, Andrus-2022}.

Yet, there remain open questions about how the perspectives and experiences of visualization designers factor into their design process and their understanding of power and politics, particularly when visualizing demographic data about marginalized populations. Ongoing work on feminist and critical epistemologies has largely been on the experiences of visualization audiences and researchers, focusing on the arrangements of power within collaborations~\cite{akbaba2023troubling} and how audiences relate to data visualizations~\cite{peck, he2024}. However, this work does not examine how visualization designers, as the creators of these artifacts perceive and account for the influence of power and related concepts such as neutrality and politics in their processes. Building on previous work led by the first author~\cite{dhawka2023we} that highlights the sociotechnical challenges of visualizing racial demographic data, we investigate the hurdles that visualization designers encounter when creating data visualizations of race and gender, how they understand politics, power, and neutrality, and how they perceive the influence of their beliefs, values, and biases on their design process. Inspired by feminist and critical work in HCI~\cite{datafem, akbaba2024entanglements, Ogbonnaya2020, bardzell-bardzell-2011}, this work aims to answer the following research questions:
\begin{itemize}
    \item \textbf{RQ1} What challenges do visualization designers encounter when designing data visualizations of protected demographic categories such as race and gender?
    \item \textbf{RQ2} How do visualization designers' understandings of politics, power, and neutrality help us understand data visualization design as socially constructed, having politics, and value-laden?
    \item \textbf{RQ3} How do visualization designers' perceptions of their positionalities, beliefs, values, and biases on their design process and the resulting data visualizations help us understand data visualization design as socially constructed, having politics, and value-laden? 
\end{itemize}

To elucidate the challenges and experiences of designers visualizing demographic data such as race and gender, we conducted semi-structured interviews with seventeen visualization practitioners working in various fields ranging from public health policy to data journalism. To our knowledge, this work is the first to study how visualization designers perceive the impacts of their positionalities on the artifacts they create, alongside their understandings of power, politics, and neutrality in their work. We offer three contributions. First, we present qualitative insights from an interview study with visualization designers on their challenges, design processes, and lived experiences. Second, we examine how visualizations of protected categories such as race and gender are subjective and socially constructed by applying the feminist concept of situated knowledges~\cite{haraway1988situated} to our participants' experiences, discussing how designers navigate tensions involving their values, identities, power, and politics in their work. Lastly, we conclude with a set of considerations for research and practice that center the experiences of visualization designers working with protected demographic data. 

\section{Related Work}
In this section, we summarize prior relevant work in HCI and information visualization with a focus on (1) critical and feminist thinking for visualization research and HCI, (2) visualizations of demographic data, and (3) practices of visualization designers. 

\subsection{Critical and Feminist Approaches in Information Visualization}
Early work by ~\citet{dork} proposed a critical approach for visualization research that centered values such as disclosure, plurality, contingency, and empowerment. Emphasizing the potential of data visualizations to influence high-stakes decision-making, Correll~\cite{Correll} argued that visualizations are not neutral and that researchers in this space should carefully consider the ethical implications of their work. Correll proposed a list of obligations for the visualization community, including visualizing hidden labor, collecting data with empathy, and challenging structures of power. Extending this call from a broader data perspective, D'Ignazio and Klein~\cite{datafem} proposed the data feminism framework, drawing on feminist epistemologies to question the role of power in data processes. Prior critical work in HCI has also examined the challenges of working with racially marginalized populations and their data, including work that adapted critical race theory for HCI~\cite{Ogbonnaya2020} and a historical breakdown of the dimensions of race and their construction in current sociotechnical systems and research practices~\cite{hanna2020towards}.

A growing body of research has also aimed to broaden current visualization practices to include concepts such as care, equity, and feminist entanglement. Schwabish and Feng's Do No Harm guide~\cite{SchwabishFeng} offers guiding principles for data practitioners that center equity particularly when visualizing demographic data. Using the lens of matters of care, Akbaba et al.~\cite{akbaba2023troubling} highlighted the tensions related to power and responsibility in visualization research collaborations between researchers and domain experts. Recent work by Akbaba et al.~\cite{akbaba2024entanglements} traced the genealogical lineage of feminist entanglement theory and illustrated a case study of applying feminist epistemologies to visualization research. In this work, Akbaba et al.~\cite{akbaba2024entanglements} introduced situated knowledges, a feminist epistemology proposed by~\citet{haraway1988situated} that emphasizes that all knowledge is situated in the perspective of the person producing said knowledge, which in turn determines what, how, and how much they can know and can thus communicate to others. We examine how visualization artifacts are shaped by the beliefs, values, politics, and positionalities of designers and frame our findings using Haraway's concept of situated knowledges~\cite{haraway1988situated}. 

\subsection{Visualizing Demographic Data}
Research on visualizing demographic data has broadly focused on best practices (such as using color and language mindfully~\cite{SchwabishFeng, Datawrapper-Race}) and open questions around equitably representing these categories~\cite{cabric2023open, dhawka2022representing}. Specifically, this work has included visualizations of gender, with ~\citet{Cabric2024} summarizing visualization practices in five academic communities, ~\citet{tovanich2022} investigating gender representation in 30 years of IEEE VIS publications, and ~\citet{elli2022visualizing} presenting an artistic visualization of sexual harassment. Although these efforts reflect growing interest in equitably visualizing demographic data, the visualization research community has yet to critically engage with how the positionalities of researchers and practitioners, who remain predominantly male and White~\cite{WEIRDCHI, oconnor2023decolonizing}, might influence their approaches in working with demographic data of groups they do not belong to. Prior work has also explored anthropographics (human-shaped visualizations) for their potential to evoke prosocial feelings from audiences in specific humanitarian contexts ~\cite{Boy, Morais2020, Morais2021}. In response to homogeneous anthropographics~\cite{Boy, Morais2020, Morais2021} with generic human shapes that obscure the demographic differences of the people being visualized, Dhawka et al.~\cite{dhawka2023we, dhawka2024} proposed diverse anthropographics---visualizations that communicate demographic diversity using physical characteristics---and provided a broad overview of the research challenges around representing racial data. Extending this work~\cite{dhawka2023we, dhawka2024}, we investigate the specific practices and values of designers producing visualizations of demographic data of race and gender, including their use of diverse anthropographics. 

\subsection{Practices of Visualization Designers}
Existing research on visualization practitioners has largely examined their design processes in attempts to bridge knowledge gaps between the academic and practitioner communities. Parsons investigated the design processes of professional data visualization designers, including their decision-making steps, strategies, and familiarity with popular design methods~\cite{parsons2021understanding}. Parsons' findings revealed that visualization designers do not follow pre-determined steps but rather engage in a reflexive and intuitive process, as a result of their ``situated knowing''~\cite{parsons2021understanding} that are often absent from idealized models of visualization workflows. Subsequent work by Parsons et al.~\cite{parsons2021fixation} also investigated how data visualization creators engage in design fixation in their process, focusing largely on creative design practices. Work by Bako et al.~\cite{bako2022understanding} has examined how visualization designers use examples in their process, particularly in data storytelling. Within the context of the COVID-19 pandemic, Zhang et al.~\cite{zhang2022visualization} investigated the experiences of visualization dashboard designers and their strategies for handling potential misinterpretation of their work. While prior research has focused on the strategies and design processes of visualization practitioners more broadly, we specifically investigate how practitioners experience the production of visualizations of race and gender, focusing on how the nature of this demographic data interacts with issues of power, politics, and neutrality.
\section{Key Conceptual Terms}
\label{sec:conceptual-terms}
We briefly describe the conceptual terms central to our analysis and their specific use in this work: power, politics, positionality, race, gender, and feminist and positivist approaches. We cite descriptions from work within HCI, including literature adjacent to visualization research, or work that has been extensively cited in the HCI literature, with the acknowledgment that all of these concepts originate from and have rich intellectual histories beyond HCI.

From D'Ignazio and Klein's data feminism framework, we describe \textbf{power} as ``the current configuration of structural privilege and structural oppression, in which some groups experience unearned advantages''~\cite[p.~24]{datafem} to study how designers experience power arrangements in their design process. Following Winner's highly-cited essay, ``Do Artifacts Have Politics?''~\cite{Winner1980-WINDAH-3}, we can generally assume consensus within the broader HCI community that the design of technology has political implications. Specifically, we use Winner's definition of \textbf{politics} to inform our conceptualization of the term  as ``arrangements of power and authority in human associations as well as the activities that take place within those arrangements''~\cite[p.~123]{Winner1980-WINDAH-3}. We use this definition to describe the personal politics of visualization designers and the web of power and politics they navigate when working with demographic data. 

~\citet{Liang2022} discussed the tensions experienced by HCI researchers when working with marginalized populations. Although our study population differs from Liang et al.'s~\cite{Liang2022}, we use their definition of \textbf{positionality} to describe the experiences and backgrounds of visualization designers that shape their unique perspectives of the world and their work. In this work, we refer to this set of beliefs, values, and biases as being part of a designer's positionality. 

Our understanding of \textbf{race} and \textbf{gender} as socially constructed categories is informed by~\citet{Ogbonnaya2020}'s work on adapting critical race theory for HCI,  ~\citet{hanna2020towards}'s work on the multidimensionality of race for algorithmic systems that often operationalize race to a single dimension (phenotype), and ~\citet{Keyes2018}'s work on how algorithmic systems conceptualize gender as binary biological categories. Our participants worked with racial data primarily derived from the US census and collected via racial self-classification within eight categories (American Indian, Asian, Black, Hispanic or Latino, Native Hawaiian, Other, Two or More, and White~\cite{censusUSA}). Additionally, the US census does not currently collect data about gender but instead asks individuals to self-identify in terms of sex, which is restricted to two categories (male and female)~\cite{censusUSA-Sex}. In contrast, our participants discussed how they visualized gender categories (such as nonbinary and transgender) beyond the binaries present in US census data.

As the most topically relevant published work on feminist theory for visualization research at the time of writing, we use the term \textbf{feminist} from Akbaba et al.'s description of feminist theory as work that ``centers the role of power and privilege''~\cite[p.~1280]{akbaba2024entanglements} in contrast with \textbf{positivist} approaches  that emphasize neutrality, objectivity, and independence between researchers and participants~\cite{duarte2016, bardzell-bardzell-2011}.

\section{Study}
We conducted seventeen semi-structured interviews with diverse individuals working in data visualization roles across the US to investigate the impact of their politics, beliefs, values, and biases alongside the challenges they experienced when designing visualizations of protected demographic categories. Our university Institutional Review Board (IRB) reviewed our study protocol and determined it to be exempt.
\subsection{Recruiting Participants}
We focused on recruiting visualization designers working with race and gender demographic data. To recruit interested individuals from underrepresented groups who we may not have reached through conventional recruitment methods, we posted recruitment messages on visualization-specific subreddits (r/visualization)\footnote{\url{https://www.reddit.com/r/visualization}}. We also recruited via the Data Visualization Society's monthly call for research opportunities\footnote{\url{https://www.datavisualizationsociety.org/}} to more broadly reach visualization designers across the US.

To be eligible to participate in this study, participants were required to reside in the US, be fluent in English, and have at least 2 years of experience designing data visualizations of race and other demographic data in a professional capacity, either as a freelancer or as part of a broader organization. Our recruitment messages included a pre-interview questionnaire to screen for participants who met the study eligibility requirements. This initial round of recruiting yielded 12 interviews with participants with some internal experience with demographic data but was missing individuals with full-time experience designing visualizations for external audiences such as the general public. Furthermore, this sample of 12 participants was heavily skewed towards individuals identifying as male and working in federal and state public policy positions. To diversify our participant sample, we reached out via email to femme-presenting designers in data journalism roles at a variety of US based news outlets. As these positions frequently involved designing visualizations of race and gender data for broad audiences, we connected with designers who had worked on data stories of race and/or gender published in 2024 using their publicly available contact information. This yielded 5 interviews with individuals in various visualization and data journalism-related positions. 

\subsection{Participants}
Our final sample of interviewees consisted of 17 participants, including 8 women and 9 men who self-identified their gender via an open-ended textbox~\cite{Spiel2019}, with professional experiences from various fields, including the music and fashion industry, federal health agencies, private data consulting companies, nonprofit organizations, and data journalism outlets. 12 of our participants identified as White and 5 participants identified as belonging to racially minoritized groups in the US. Participants were also geographically spread across the US and represented several political and educational backgrounds. The majority of our interviewees had graduate degrees and 2 participants held teaching positions at higher education institutions. To ensure the anonymity of our participants and following suggested best practices in HCI~\cite{SchwabishFeng, Spiel2019, Chen2023}, we are choosing not to include additional disaggregated demographic data.
% Within our sample we noted 4 data journalists, 2 freelancers, and 5 individuals with prior or current experience working in state and government agencies

\subsection{Procedure}
We used a semi-structured interview protocol (see appendix \ref{appendix:protocol}) with guiding questions to prompt discussion around the challenges and politics of visualization designers working with protected demographic data. We conducted remote interviews of up to an hour via Zoom conferencing software. The first half of the interview focused on the background of the designer, their work experience, tools, and design practices. The second half was guided broadly by the themes of identity, power, neutrality, and the impact of the designer's positionality on their process. At the end of the interview, participants were offered a \$20 (USD) gift card as compensation for their time. 1 participant refused the gift card due to ethical requirements at their workplace. Lastly, we used the Zoom-generated audio transcripts from the interviews and manually fixed  errors in the transcribed text before analysis.

\subsection{Analysis}
To analyze the qualitative data from our interview transcripts, we performed reflexive thematic analysis to extract themes related to participants' understanding of neutrality, power, politics, and positionality in their design processes~\cite{clarke2017thematic}. In the open coding stage, the first author iteratively analyzed the transcripts, generating codes until they reached a saturation point. The first author also met weekly with the senior author to discuss the outcomes from the open coding round or instances where the first author felt that they might be biased in the coding process. The first author then synthesized these codes into higher-level themes based on their recurrence in the data and similarities or differences between transcripts. 

\section{Findings}
Designing visualizations of protected demographic data such as race and gender often involves practitioners navigating tensions at various points in the design process. Below, we describe the key themes and tensions experienced by our participants as visualization designers, which we conceptualized from our analysis and are ordered according to our initial research questions.

\subsection{Design and Data Challenges} 
\label{sec:design-data-challenges}
Our participants encountered several hurdles throughout their design processes, noting that these challenges were often linked to the nature of demographic data or a lack of control over the data collection process. From these responses, we observed that participants were acutely aware of existing issues and stereotypical assumptions inherent in the  data collection process. 
\subsubsection{Demographic Categories Are Socially Constructed}
\label{sec:demographic-data}
The US government collects population-level demographic data through the census and issues in census demographic categories tend to permeate datasets derived from this bigger data corpus~\cite{censusUSA}. Participants noted that their datasets largely consisted of binary gender categories and that they frequently had little control over the data collection process. For instance, P14 pointed out how gender categories were limited to two main categories (male and female). As a result, US census data often had issues with ``conceptual missing data, because people who do not identify in the binary fundamentally, are not captured'' (P14). Likewise, P12 observed how current gender demographic categories represent how individuals are classified into certain categories at a given point in time and revealing one's gender identity might put them at risk of harm and how individuals from marginalized populations may have historical reservations about sharing their data with the US government: 
\begin{quote}
    The gender data is really a snapshot in time, like how somebody identifies now could change in the future. And that a lot of these people, especially historically, if [...] the data set goes back to like the 1950s, it wasn't safe to be publicly out as non-binary [or] transgender.
\end{quote}
Additionally, personal identities may be intricately linked with demographic categories but may not necessarily align with the categories that individuals are classified in. Reflecting on their experiences of (dis)ability and the arbitrariness of demographic classification, P12 shared how the process of realizing the fluidity of their own identities impacted their design process: 
\begin{quote}
    So I was once very able-bodied and now I'm disabled, and knowing that those identities can shift and change over time and are not stagnant, I think, probably influences me a lot. [...] And then kind of knowing that on a society level, we're putting everybody into boxes so that we can tell these larger stories, and knowing that those boxes are very iffy and messy.
\end{quote}

Our participants pointed out how US census demographic categories are socially constructed---as these definitions tend to fluctuate based on several factors, including societal perceptions of demographic groups (gender being seen as more fluid than in the past, racial categories being expanded to include more detailed breakdowns). The US government enforces these categories by tying demographic representation in governmental datasets to political representation through the drawing of election districts to elect public officials and demographic data-driven decision-making for the allocation of public resources~\cite{pine2015politics, forest2005changing}. 

However, demographic classification also occurs beyond the census and has different implications for the categorized groups. For instance, companies may collect other forms of demographic data using arbitrary measures for groups not necessarily captured through the census~\cite{feinberg2022everyday}. Focusing on the category of the ``middle American'', P10 shared their experience with arbitrary categories as a data consultant and described how these categories are then used in decision-making processes for marketing: 
\begin{quote}
    So this company, [redacted], wants to focus on the middle American consumer. They have no data set, no database. [...] How do you decide what's a middle American? They have to live in Iowa in the middle of the country. [...] That's a pretty good definition of middle American. Is it somebody who goes to church, but not a mosque? It's somebody who drives  [a] pickup truck, but not [a] Prius? How do we know? 
\end{quote}
The arbitrary demographic category of the ``middle American'' is not collected through the census but is used as a proxy for specific demographic data of race and/or gender. Thus, data that serve as proxies for race and gender can also have important political implications for the groups being categorized. Our findings highlight our participants' awareness of issues in datasets of demographic categories which consequently translated in their design process and their best practices. 
\subsubsection{Strategies to Visualize Race and Gender} 
\label{sec:strategies}
Participants underscored the benefits of visualizing demographic data to highlight trends or reveal potential gender or racial inequity in their datasets. We describe participants strategies for mitigating the challenges of working with race and gender datasets. 

Describing their experiences reporting on a data story about gender in the music industry, P12 pointed out that ``a lot of times it's hard to show like the absence of.. what's not there in the data'' and ``you can't like visually see the absence of transgender and non-binary folks because they're not there, to begin with.'' Our participant chose to explicitly visualize the absence of women of color in the dataset to bring attention both to the missing data and more systemic issues of representation in the music industry by showing blanks instead:
\begin{quote}
    [A]nother example of trying to figure out how to show that absence is there's a part where we [...] highlight just some pictures of albums that have charted by women of color and appeared on the charts of women of color and we specifically left a blank space for Asian women because there was no representation there. But we wanted to show that gap.
\end{quote}
Meanwhile, participants with an academic background shared their experiences designing visualizations of racial data using \textbf{anthropographics} (human-shaped visualizations) and were also familiar with the issues inherent in those kinds of visualizations. One participant, P14, who had extensive experience designing visualizations of race and gender, was also an advocate for equitably visualizing data about underrepresented populations. P14 focused on the pros and cons of using certain visualization designs, such as standard unit charts over homogeneous anthropographics.
\begin{quote}
    You know, icons are great in visualization because people can kind of see themselves or see the individual as opposed to an abstract shape, like a bar or a dot. But, on the other hand, there's just so many ways to misrepresent people or to [...] hide people or hide their experience.
\end{quote}
P14 also mentioned \textbf{the potential of misrepresenting the experiences of the people in the data}, highlighting that equitably representing people remains a major visualization challenge. Further reflecting on the challenges of anthropographics, P14 brought up a question that they were working on, asking ``How diverse do we make the icons versus the accurate representation of the data?'' Our participants' thoughts on anthropographics highlight the challenge of \textbf{accurately representing the people in a given dataset} without necessarily revealing their identifying demographic details via diverse anthropographics. 

Extending this critique, another participant, P15, who specialized in data stories that use human-shaped icons and cartoons, observed ``there is no shared upon agreement that, this color means this or this icon means that.'' Similarly, we also noted \textbf{tensions around creating visualizations of data about demographic communities that the designer did not belong to}. Specifically, P15 reflected on being an \textbf{outsider and not having the lived experiences of the communities they are visualizing} when describing their design process of visualizing racial inequity in the US:
\begin{quote}
    I'm not a Black American, so I don't know what things are like. If someone said ``hey I don't like to be depicted that way'', […] I don't have the sensitivities or the life experience [to disagree]. But when I make [race redacted] characters, it's a lot easier to say  ``No but this is why.'' 
\end{quote}
Here, P15 pointed out how \textbf{membership} within the demographic group they were visualizing also allowed them access to knowledge and lived experiences that they could accurately depict. Further discussing the challenges of anthropographics, P15 described how they used ``panels that illustrate [people] living their day-to-day life that has nothing to do with their race.'' P15 also questioned whether focusing on visualizing racial demographic data might distract from their goal of humanizing the people in the data: 
\begin{quote}
    I think when we talk about race in data visualizations or media, a lot of times in communicating [...], we just focus on that aspect of identity. But when you and I go through our everyday lives, we're not like [...] walking down the street thinking like I'm [race redacted]. I go through just thinking like any other human being [...]\end{quote}
P15's comment revealed tensions between assigned racial identity and personal identities of individuals, and how, when visualizing data about minoritized groups, designers might prioritize representing race over other aspects of identity that might more fully convey the humanity of the people being represented.

Overall, we observed tensions when participants expressed wanting to emphasize the experiences of the humans being visualized for their audiences but struggled with missing data about marginalized groups and separating a specific demographic category from individual identity, which may have design implications for visualizing absent data and designing diverse anthropographics.

\subsection{Tensions Around Making Comprehensible and Public Data Visualizations}
\label{sec:comprehensible-public-datavis}
Our participants worked in various professional fields and thus had experience designing for diverse audiences. In this theme, the question of demographics was present in two ways---first, the designers took into account the demographic composition of their audiences, and second, issues of comprehensibility, openness, and transparency took on a heightened status when it came to visualizing demographic data. In particular, participants mentioned making their visualizations comprehensible and changing their design processes to meet audience needs. Participant definitions of comprehensibility centered around how visualizations can be made accessible (not restricted by one's abilities) and public (available to anyone to view and interact with). \textbf{Importantly, we note that visualization designers' descriptions of comprehensibility primarily focused on making their visualizations understandable to diverse audiences with varying levels of data and visualization literacies.}
% accessibility in HCI has generally been understood as centering the experiences of people with disabilities~\cite{10.1145/3411764.3445412}.} 

\subsubsection{Designing Comprehensible Visualizations for Specific Audiences}
\label{sec:comprehensible-vis}
Participants made several assumptions about the demographic composition of their audiences and prioritized distinct visualization approaches. In their data journalism role, P17 noted how their awareness of their target audiences being a niche and exclusive group of domain experts influenced the data stories they worked on, noting that ``we want stories that will do well with those [redacted] clients [...] [who] tend to skew richer, White, top of the line people.''
% \begin{quote}
%     Yeah, at this point I definitely have a strong sense of who the audience is, especially around pitching stories. [...] And so usually we want stories that will do well with those [redacted] clients [...] [who] tend to skew richer, White, top of the line people. 
% \end{quote}

Participants then used these informed assumptions to choose the appropriate visualizations for each audience. P6, who worked with racial demographic data in a public policy and health role, described the comfort of their perceived audience with certain types of charts as ``pretty comfortable seeing maps and knowing where counties are when they're looking at where to put in programs or new hospitals or new facilities.'' Similarly, P4 described how their process changed depending on how they perceived the audience's data visualization literacy, noting  ``if I'm designing for an audience, that's not data literate or digital literate, I definitely simplify it.''
% \begin{quote}
%     And users are pretty comfortable, especially in [redacted location] and the target audience for this, they're pretty comfortable seeing maps and knowing where counties are when they're looking at where to put in programs or new hospitals or new facilities. 
% \end{quote}

Participants often mentioned the motivation behind their strategies to connect with or retain the attention of their audiences. P15 briefly mentioned wanting to individually impact their audience, particularly people who did not have a college education and were neglected by news outlets, emphasizing ``I want people who don't typically read news media, who need to feel like I'm talking to them, [to read this].'' Similarly, P8, who worked in a data role at an educational institution, described how they provided additional context to their visualizations for specific audiences: ``When I design something for our teachers, I want them to be able to use it whenever and wherever. So I really have to build in a lot more of that context.'' Furthermore, participants described strategies to make their designs more comprehensible to audiences. In particular, as an outspoken advocate for data ethics at their workplace, P5 shared their understanding of accessibility, as ``making it accessible for everybody, not just accessible for certain people, but another dimension of accessibility is the ability to comprehend.'' 

% \begin{quote}
%     And so I'm going to design for you and write for you, and not [...] like other journalists, not to try to impress other smart college-educated people, even though those people enjoy it, because, for them, it's a ``go look at this fun thing that's about something I care about.'' But I want people who don't typically read news media, who need to feel like I'm talking to them, [to read this]. 
% \end{quote}
% }

Participants also discussed their considerations of design decisions around the complexity and aesthetics of the final design among the factors that could make a visualization more comprehensible to their audiences. Participants working as data journalists remarked on how perfectly designed visualizations of race and gender can feel disconnected from real life and be less accessible to their audiences. Some participants often opted to leave some imperfections in their designs so audiences could relate to their visualizations. P12 mentioned this feeling of disconnect in complex visualizations and how they tried to make their visualizations feel more human for their audiences through imperfections: 
\begin{quote}
    So when you get a product, and it's completely buttoned up, it can feel sterile or that you are not a part of it. [...] And I want to make sure that I'm passing on that humanity to anybody who interacts with my pieces [...] a level of [...] unrefinement that also feels a little bit human, too. [...] So seeing a little bit of the imperfections... I've [...] learned to enjoy that because it [conveys] this is a little bit of a scrappy organization where we're human. 
\end{quote}

While, in general, our participants were designing for a variety of audiences, their strategies emphasized their assumptions about audience visualization literacy and the motivations behind their process, particularly wanting their visualizations widely comprehensible. 

\subsubsection{Downsides of Public Data Visualizations}
\label{sec:downsides-accesible-vis}
Participants had distinct approaches when creating visualizations that were shared internally (within their workplaces) or externally (for the general public). Specifically, designers frequently had to include additional context, provide more support for varying levels of data visualization literacy, protect the privacy of the individuals being represented, and reduce the risks of their work being misinterpreted when creating external visualizations. On the other hand, creating visualizations meant for internal use (within their workplaces) meant they could reasonably assume that their audience had sufficient context about the dataset being visualized and the designer would be available to answer audience questions. 

Speculating on the differences in their process when designing for an external audience including the general public, P1, who previously worked in a research data role at a small East Asian country before moving to the US, described potential unintended side effects of releasing internal visualizations to the general public of said East Asian country: 
\begin{quote}
A lot of our work was not released to the public unless it's deemed sanitized enough to be featured [in] a press release. By putting this information out into the world, is it going to drive people's attention to very destructive discourse on the differences [between them]? 
\end{quote}
Although P1 often had to consider whether publicly sharing their work could negatively impact public opinion, other participants often received support from their colleagues and employers before sharing their work with external audiences. In particular, P11, working in a public health policy role, emphasized the additional checks and balances in their workplace infrastructure when creating and before releasing visualizations meant for the general public: 
\begin{quote}
    Anything that is going beyond our Board of Health or internal sharing capacity goes through our communications team because they're responsible for everything that goes out to the public. So they want to verify what's going out. One, they have talking points prepared for if they get questions on and make sure that we didn't make a silly spelling error or formatting error, or something like that.
\end{quote}
In this case, designers had to follow certain best practices along governmental and workplace guidelines around sharing protected demographic data with the public. 

When creating visualizations for external audiences, participants also mentioned wanting to support individuals in exploring the data on their own. Participants did so by broadly \textbf{resisting specific interpretations of the data} and using best practices to make their visualizations comprehensible to a broad audience. In the context of the COVID-19 pandemic\footnote{COVID-19 featured prominently in our participant responses as the pandemic was a period of rapid creation and proliferation of visualizations of demographic data, as~\citet{lee2021viral} demonstrate.}, P11 shared the challenges they experienced while creating visualizations of race and gender that would allow their audience (the general public and audiences holding political office) to explore the data: ``I had [audiences redacted] saying, `Well just give us the raw data and we'll do the analysis because we don't trust you.' In this instance, P11 was also working in a politically charged role during the COVID-19 pandemic and often experienced distrust from their audiences who wanted access to the demographic data to do their analysis and visualizations. However, making visualizations and datasets of protected demographic categories publicly available also runs the risk of exposing these artifacts to misinterpretation or misinformation~\cite{lee2021viral}. 

Thus, our participants also worried about the unintended consequences of releasing visualizations without context or interpretation, which in turn can be used to spread misinformation about the groups being visualized. Reflecting on rampant misinformation in the US, P13 was mindful of their design process as a data journalist trying to avoid creating visualizations that might be misinterpreted: 
\begin{quote}
    We have to think a little bit about the audience that's going to either be willfully misinterpreting it and how to make sure that our stuff can't be used to fuel conspiracies [...] But also that doesn't mean we should fully give up on trying to ground our work in something that feels complete in the sense of where we're at, at this point in history, and also can't be super easily discredited.
\end{quote}
 Similarly, P11 further described their role as someone working with sensitive demographic and health data in a government agency and the potential of their work being misappropriated: ``So we're advocates for the data on all fronts. Including how others are trying to use it. That's why we're always really careful with what we share in how we share it.'' The approaches described by P13 and P11 mirror those of several other participants who also emphasized how they resisted specific interpretations of the data. Overall, participants stressed the role of the designer in communicating data that reflects the truth, bringing up tensions around neutrality that we discuss in \S\ref{sec:tensions-neutrality}. 
% P9 briefly mentioned their experience designing visualizations in the current post-truth climate: 
% \begin{quote}
%     So in this political environment, it is very difficult to know what is factually correct and what it's very easier, yet very difficult to convey what is factually correct and what's factually incorrect. 
% \end{quote}

% \subsubsection{Visualizations Perpetuating Information Inequality}
Reflecting on the comprehensibility of their visualizations, our participants observed that public data visualizations were often complex and inaccessible to both expert and general audiences. For instance, when discussing COVID-19 data visualizations, P11 remarked that data visualizations designed by government agencies (in this context, the US Center for Disease Control or CDC) for the general public during the pandemic were frequently inaccessible and not easily understandable, even to domain experts like themselves:
\begin{quote}
    If you look at any CDC graphic, and this is not trashing the CDC, but their graphics are terrible, and I, as a master's level educated epidemiologist, have a hard time interpreting some of their graphs. 
\end{quote}
Here, we note that P11's experience reflects findings, albeit from the perspective of a visualization designer, by Lee et al.~\cite{lee2021viral} who investigated the lack of easily understood public data visualizations created during the COVID-19 pandemic. 

The complexity of public data visualizations often reinforces existing systemic inequalities around data visualization literacy and information access, which are also related to the use of these visualizations in the spread of misinformation~\cite{lee2021viral, lisnic2023misleading, lisnic2024}. As a data journalist now working in an academic position, one participant, P15, described the phenomenon of \textbf{information inequality} where the default audience of data visualizations often belong to certain demographic groups that skew wealthy and educated, neglecting lower-income individuals who may be assumed to be less data literate. Here, P15 described how they were part of the highly educated target audience of data journalism in news media but were also designing data visualizations for said audience:
\begin{quote}
    I think one realization I had was news media [...] has become highly educated, highly privileged. If you look at the median incomes of people who read the news versus people who don't,[...] there's a huge gap. [...] a lot of times, people who are in these [data journalism] circles, who are constantly writing about these things don't really see it, because their entire lives have been around these worlds. But it's important for me to say [...] I know how to talk like you [as a journalist], but I refuse to talk like you. Cause then that's not doing our job.
\end{quote}
However, with their awareness of existing systemic inequities in information access, P15 reflected on how data journalists may unintentionally perpetuate information inequality through complex data visualizations. Thus, P15 affirmed their commitment to creating comprehensible open visualizations to fight information inequality. Similarly, P13 observed ``it's really hard to break out of, to your point about audiences... maybe we're doing this for a very small academic group'', reflecting on their awareness that they were designing for a small academic group, despite working at an outlet meant for a broader audience. 

Although our participants prioritized comprehensibility in their process, they often experienced challenges that made their work inaccessible to broader, external audiences. In particular, we found that our participants were aware of the potential of their work being inaccessible to audiences with varying levels of data visualization literacy and being misappropriated for misinformation. Lastly, our participants briefly discussed the phenomenon of information inequality that has broader implications for making comprehensible public data visualizations that are resistant to being misappropriated for misinformation. 
\subsection{Tensions Between Neutrality, Objectivity, and Other Values}
\label{sec:tensions-neutrality}
Neutrality is often seen as a desirable, yet challenging value when creating data visualizations. Here, we discuss how participants described their understanding of neutrality as designers, illustrating the ambiguity of the term and how the differences in perspectives translated into their design practices. We note that we did not provide participants with a description of the term ``neutrality'' and instead asked interviewees about their understanding of the term.

\subsubsection{Participant Understandings of Neutrality}
\label{sec:descriptions-neutrality}
Participants shared various personal definitions of neutrality and we highlight a smaller sample of these responses for brevity. P14 defined neutrality as ``primarily analytical and statistically based, acknowledging sort of more and more that we have fundamental and longstanding data collection issues that we have yet to resolve'', hinting at broader systemic issues in demographic data collection. In contrast, P9 described \textbf{using the data to verify the neutrality of their work} and using their design to communicate that they are neutral: 
\begin{quote}
    I just have power in my design, and through my design, I should be conveying what is certainly correct and factually accurate without having any biases [...] that there is a certain way of communicating this rightly and accurately through my design rather than imposing anything. 
\end{quote}
Similarly, P10 understood neutrality as ``stepping out of the picture, and not imposing any viewpoint on the data and stuff'' to provide an unbiased view to their audience.  
 
Participants also acknowledged \textbf{neutrality as a challenging concept to define and follow}. For instance, P6 reflected on how they could implement safeguards and follow guidelines to be neutral and still not create neutral artifacts: 
\begin{quote}
    I'm not saying there is no such thing as an objective truth, but I don't think any single person has access to it, let alone me. I think if I use other heuristics, I might be getting more towards something where my design is neutral. But that's still not the case. Even if I have others reflect on my work and provide feedback, the others that I'm working with, the others that I encounter are disproportionately likely to be like me, sociodemographically right? So that's not going to get to that neutral spot.
\end{quote}
P6's response highlights the challenges of practicing and evaluating neutrality as a design value in their work.

Participants also provided differing perspectives on the stereotype of the designer as a neutral communicator. Specifically, participants emphasized their role as visualization experts who were providing recommendations to their audiences and were careful to not prescribe or influence the outcomes of the visualizations. As someone working for a government agency, P11 described the nuances of \textbf{recommending but not prescribing to their audiences}: ``we try to remain neutral as best we can. But of course there are certain things that we are advocates for but we always start at a place of recommendation right?''
A key theme corresponding to neutrality was \textbf{resisting specific interpretations of the data}. P7 shared how they held back from conveying any form of judgment in their visualizations and how adding typical visual elements such as annotations can result in audience backlash to their designs:
\begin{quote}
    I try to leave out statements where I conveyed judgment. I've definitely in the past learned that through [the] school of hard knocks, when I've added arrows to my data visualizations like ``Look right here, here's the problem.'' [...] I've definitely gotten push-back of people arguing like ``nope, that's your opinion on it.''
\end{quote}
From our participant experiences, we observed that the practice of recommending but not prescribing aligns with popular visualization best practices but can also make visualizations more vulnerable to misinformation~\cite{lisnic2023misleading, lisnic2024}. Although these responses focused on defining neutrality as a concept, other participants provided a range of definitions that supplemented or contrasted neutrality with objectivity as another design value, which we discuss next.

\subsubsection{Distinguishing Between Objectivity and Neutrality}
\label{sec:differences-neutrality-objectivity}
Some participants emphasized valuing objectivity over neutrality in their design. Importantly, our semi-structured interview protocol did not include questions about ``objectivity'' but rather, this concept emerged organically in our conversations with our participants. P7 described presenting the ``numbers'' as being objective and neutral but sharing their opinions as being subjective: 
\begin{quote}
     When I start [...] explaining why, using my opinions [...] I don't have data for that. And that's when I feel like I'm starting to get out of neutrality and starting to be subjective and starting to push my own opinions in the analysis.
\end{quote}
From P7's description, \textbf{resisting specific interpretations of the data} is an example of being objective and neutral whereas adding a personal element, such as an opinion, diverges from neutrality. On the other hand, P5, who worked for a US government agency, explicitly identified objectivity as a design value, which they used to describe their role as a visualization designer: ``I think that my job as a visualizer is not necessarily to be neutral. It's to be objective and this is a conversation that I think a lot of institutions need to have, especially the media.'' Meanwhile, P16 valued objectivity over neutrality which they associated with inaction:
\begin{quote}
    I don't think that's a term that registers for me and my work. I think that neutrality is very different than objectivity. But I do think that comes more in my mind before neutrality, feels like I associate it with inaction, like not doing anything.
\end{quote} 
Thus, some participants offered objectivity as an alternative to neutrality while also questioning the benefits of prioritizing neutrality in their process.  

\subsubsection{Value-Laden Approaches and Their Downsides}
\label{sec:value-laden-approaches}
Some participants prioritized other values besides neutrality in their process. P6 expressed wanting to communicate minoritized perspectives to humanize marginalized populations, highlighting how a value-laden visualization approach can help disrupt existing power arrangements by shifting the focus away from dominant demographic groups towards lesser-known perspectives: 
\begin{quote}
    So I question sometimes the value of neutrality. I think sometimes I want to be looking for those unheard perspectives, and I want to be emphasizing those. [...] If I can help, if I can focus on and bias myself towards a minority or oppressed group [...] and bring them more value relative to people that are already in power or privileged, [that] might not be neutral.
\end{quote}
Following a similar approach, P12 understood neutrality as a default concept usually associated with reporting the facts and consequently removing the human element in visualization and data journalism.  
\begin{quote}
    What society often determines as neutrality is actually just [...] the White cis male version of events. [...] And oftentimes I think that it has kind of done a lot more harm to the industry than it has done good. Just because we've taken the people and the humanity out of news. 
\end{quote}
% P12 described neutrality within the context of their work reporting on the rapid turnover in President Trump's cabinet during his presidential term and the online harassment that they received over this data story.
P12 described \textbf{neutrality as visualizing dominant perspectives without acknowledging broader systemic inequities that impact vulnerable and marginalized populations}. By simply reporting the facts without additional historical and social context, a neutral position can have the opposite effect of humanizing the people behind the data. P12's response resonates with critical work in HCI, such as work on seemingly neutral platforms such as Wikipedia~\cite{menking2021wp}, that situates neutral approaches as being implicitly value-laden. 

However, participants noted that a value-laden stance had potential downsides for their work as designers. Another data journalist, P13, claimed neither neutrality nor objectivity as design values but shared personal observations about providing a complete picture of the data to audiences: 
\begin{quote}
    So I am of this dual belief that there is no such thing as pure objectivity. There is no such thing as pure neutrality. We're all bringing our perspectives into it. But also that doesn't mean we should fully give up on trying to ground our work in something that feels complete in the sense of where we're at, at this point in history, and also can't be super easily discredited. I think there's credible work being done by activists or political parties, or whatever. But when that's so clearly pushing something, it's not going to be valued or trusted as a source of information in the same way. And so for me, it's like, how do you balance where you come from in your own perspective?
\end{quote}

P13 acknowledged that their process is situated in their personal experiences while being aware that \textbf{value-laden approaches that advocate for specific interpretations are often dismissed as being less trustworthy than seemingly neutral approaches}. In summary, our participant responses provided varied insight into how designers might conceptualize neutrality and objectivity alongside other design values. 
\subsection{Navigating Politics and Power}
\label{sec:politics-power}
Our findings exemplify the agency of visualization designers and their options to refuse work that may have harmful consequences on marginalized populations. For instance, designers can refuse to create demographic data visualizations they disagree with or find strategies to disengage with the work they are assigned. In this section, we interpret participant responses to interview questions on power and politics in their design processes. 

\subsubsection{Relational Understandings of Power}
\label{sec:relative-power}
Participants understood power as a relative concept in their working relationships and in their design. For brevity, we highlight select participant definitions of power. For instance, P16 described the power that comes with seniority at their workplace: ``I would say, I have a lot of control. [...] I think that I can kind of propose anything I want. I'm the most senior datavis person at the company.'' On a related note, P13 described checks and balances in the data journalism publication pipeline that restricted their power: 
\begin{quote}
    In this one, my boss gives me pretty much complete editorial freedom. But he gives advice when I ask for it, which is all the time. [...] Typically, I get to have whatever I want. And there's edits. And there's a process to make sure we're being responsible. 
\end{quote}
Although we only highlight select quotes, participants primarily shared how they had power relative to their colleagues in their workplaces and power in the levels of creative freedom they were allowed in their designs. We suspect that participant responses were limited since we did not provide a definition or examples of power in our interviews but rather relied on participant understanding and awareness of the concept.

\subsubsection{Political Impact of Their Work}
\label{sec:political-impact}
Several participants described reflecting on power and politics when working on high-stakes visualizations that were directly used for decision-making. For instance, P8 recalled their experience when they worked directly under someone holding political office and were designing demographic data visualizations during the COVID-19 pandemic, noting how ``there is politics involved at that high level''. Since P8 created visualizations that were used for decision-making around public health measures and the allocation of financial resources, their work had direct political implications for the demographic groups being represented, which they described as having ``a political slant''.
% \begin{quote}
%     I think the [redacted audience] did a really good job of telling us or asking us to just give [them] our opinions. Obviously [they] were appointed by [redacted], and there is politics involved at that high level, union politics, things like that. But I think [they] did a really good job of genuinely asking us for our opinions and thoughts. But we definitely felt like the questions that [they] asked us to research week to week add, I wouldn't say, a political motivation, but like a political slant to it. 
% \end{quote}
Working in the context of federal medical policy, P16 also shared the potential political implications and responsibilities associated with their work, noting how misrepresenting a state's demographic data can result in ``a lot of possible political fallout that can happen from it, if for various reasons...'' and that their role ``is trying to ensure that the data is understandable and not misrepresented.''
% \begin{quote}
%     But I think for [redacted company], a big concern, is if we misrepresent a state’s [demographic] data because a lot of this is about how is the state doing in their delivery of healthcare and their provision of healthcare [...] People don't take it out of context, like there's this project, there's a lot of possible political fallout that can happen from it, if for various reasons... So anyway, I think that's kind of mostly my central value is trying to ensure that the data is understandable and not misrepresented. 
% \end{quote} 
% Here, P16 was acutely aware of the risks of creating misleading visualizations and reiterated their role in accurately communicating data in an accessible manner. 

Participants working in public policy or adjacent positions reflected on how their visualizations of demographic data were used in decision-making that had direct political implications, either for the groups being represented or their employers. In these instances, we observed how the process of visualizing protected demographic data can have implicit politics that are often not addressed by existing visualization workflows or knowledge of designer practices. 

\subsubsection{Working in a Politically Aligned Environment}
\label{sec:work-environment}
Most participants described being politically aligned with their employers and working in a psychologically safe environment where they could express their opinions. P16 contextualized their experiences working in a government agency, sharing ``not that there aren't flaws with government agencies, but I do think [...] the idea is in a direction that I feel is very palatable for me politically.'' Likewise, P8 described how they felt more politically aligned with their current workplace than their previous roles:
\begin{quote}
    Whereas here in [redacted] there's always politics, because that's just public government life. But [...] everyone's on the same stage [...] when we hire people, we really look for people with a strong anti-racist commitment and beliefs and things like that. And it's something that's a little bit, maybe more self selecting. So I feel like I don't feel the politics too much here. 
\end{quote}

Other participants described their experiences disagreeing with the visualizations they were tasked to create for particularly political topics, despite being overall aligned with their employer's politics. Particularly, P13 described how visualizing certain issues had more political implications than others, explicitly describing a data story about visualizing the rise in anti-trans healthcare-related bills in the US. P13 noted that their employer did not require additional disclaimers in data stories related to abortion, which is also a politically fraught topic in the current US political climate in 2024. Meanwhile, when visualizing anti-trans bills, P13's supervisors required additional context and disclaimers but de-prioritized the data visualizations. 
\begin{quote}
    It felt like they were insistent that we had a lot more nuance in a way that... we do abortion stories all the time, and we don't have to write 3 paragraphs about abortion not being safe or people regretting abortion. That's not a thing that we have to do. And yet, whenever we talk about anything like... this story itself is just a round up of the bills, and the fact that we had to do so much... and from a design perspective, really push down the chart. But all this nuance was really unusual in my experience and really frustrating, and it was something that I brought to higher ups. And I mean, ultimately we fought a lot. They wanted way more in there.
\end{quote}
 
As a senior data journalist, P13 strongly advocated against these requests. On the other hand, participants holding minoritized identities and in junior positions felt less comfortable expressing disagreement with the visualizations they were tasked with creating, especially when their work had indirect political impacts. As a junior data journalist, P17 described disagreeing with their workplace on selectively covering the war in Gaza in 2024: 
\begin{quote}
    For example, a thought that I've had working on Gaza coverage is our organization and many Western media organizations are very like, ``you have to be neutral'', and we have to [be]. But then, when you're working on the stories, you see that we don't trust Gaza data. Why don't we trust Gaza data then? But we trust Israeli data? So it's […] this lens of objectivity and neutrality that we're claiming to have, does that really exist then?
\end{quote}
As they speculated on why data about war casualties from Palestinian authorities is seen as less valid than data from Israeli authorities, P17 observed how \textbf{the language of neutrality and trustworthiness of data can be re-appropriated to silence marginalized voices}. Although the majority of our participants worked in environments where issues of power and politics were less prominent for them, participants working on explicitly political visualizations were more aware of their (lack of) power and the politics involved in advocating for their work.  
\subsection{Identities of the Designer}
\label{sec:identities}
In response to the interview question ``How would you describe the identities that are most important to you and your politics?'', participants focused on deeply personal aspects of their identities, observing that identities are fluid and can be tied to their experiences of marginalization if linked to the demographic categories they are classified into. 

\subsubsection{Identities Are Deeply Personal and Can Be Politicized}
\label{sec:personal-identity}
% P15 reflected on the intersections of their identities and demographic categories, highlighting how experiences of marginalization may be linked to identity: 
% \begin{quote}
% Those things are just biographical facts, not my identity, but those biographical facts matter to other people. [...] I don't think of them as identities that are important, but I think of them as identities that were made important because someone decided that we have to be treated differently, or you have to be treated differently, based on that identity.
% \end{quote}
As an individual with a minoritized racial identity, P15 reflected on how individuals from marginalized demographic groups may consider the corresponding identities as less important and described how their identities influence their approach to visualizing racial demographic data: 
\begin{quote}
    A lot of times, I think the really important identities in the stories that I cover are things that [are] not a super important part of your identity. [...] You didn't pick it but it ends up being important because the world we live in has shaped that and made that important, or made that either important in a good or bad way.
\end{quote} 

For some participants, their marginalized identities directly influenced their experiences designing visualizations. Particularly, as a data journalist, P17 maintained an online presence on the X platform (previously Twitter) and shared how they were frequently harassed for working on visualizations of race. Discussing Western media representation of the ongoing war in Gaza in 2024, they further described the tensions between their identities and their work being politicized against them and being perceived as less neutral when advocating against unfair media representation of marginalized groups: 
\begin{quote}
[R]egardless, at the end of the day, anyone can look at me and make whatever assumptions they want about my [data] story based off me, and, my culture, religion, whatever it is. And so [neutrality] doesn't exist. Like people can make whatever assumptions they want about the data, whatever they're looking at, and they can make whatever assumptions about my political beliefs and stuff. 
\end{quote}
On the other hand, P12 shared their experiences working as a queer data journalist in the aftermath of the deadly 2016 Pulse nightclub mass shooting in Orlando, Florida~\footnote{\url{https://www.npr.org/2016/06/16/482322488/orlando-shooting-what-happened-update}}: 
\begin{quote}
I was in a [...] traditional journalism role, and actually [...], like much of the country, [I] hadn't grappled with what identity meant intrinsically, or how my human persona differed from my journalism persona. It was just very much like, I'm a journalist. I'm gonna shut all the walls off and do what I'm supposed to do when there's breaking news. And so I made this map and only later has it come back to me and [...] haunted me of just what an awful task that was as a queer person covering this, pointing out room by room, a slaughter that happened.
\end{quote}
Describing the challenge of having to step into the role of the ``neutral'' journalist and visualization designer, P12 reflected on the lack of care from their previous workplace when they were tasked with creating a visual representation of the mass shooting without much consideration about their psychological safety as a member of the affected demographic group.

Participant identities were deeply personal and had varying levels of impact on their design processes. We found complex interactions between participants with minoritized identities and their visualization approaches, noting that participants had the doubly challenging job of maintaining neutrality while working on troubling visualizations with direct political impacts.

\subsubsection{How to Do Better as a Designer?}
\label{sec:doing-better}
In response to how their personal beliefs, values, biases, or politics influence their design process, participants discussed how they tried to engage in \textbf{allyship} with marginalized populations by being mindful of and mitigating the influence of these factors. P3 reflected on their design approach and their biases: 
\begin{quote}
    So it's more important to me to examine what my biases are coming into it and what the biases of the data itself are that I'm trying to use and how that impacts what the story is and how it can be told. [...] And so I spend a lot of time on that, and making sure that I am being aware of the challenges that I face. 
\end{quote}

% P6 on being aware:
% \begin{quote}
%     So I try to check that, one of the things when collaborating, especially that comes up with identity, is keeping in mind that white male thing, but then also keeping in mind the Adhd plus autism. 10. For like info dumping. So I try to be really, really cognizant of that, especially when talking to women, so it doesn't feel like I'm mansplaining because info dumping and mansplaining are sometimes the same thing right? And I can get so excited about a topic that it might feel like, I'm mansplaining, really like I'm not presuming the other person doesn't know what I'm saying. I'm just excited and talking. and that can happen a lot. So I try to keep that in mind with collaboration. 
% \end{quote}
% P8 on their biases influencing their design process:
% \begin{quote}
%     I know they do. There's some ways that I think I really work to, work against any inherent biases, and I'm sure there's still some biases that slip through unnoticed. 
% \end{quote}
Similarly, P15 highlighted how they tried to be attuned to broader positivist assumptions about data reflecting the truth:``
I'm very attuned to the assumptions that often go into the ways that we talk about data and information. And that I think we need to be more attuned to as well.'' On the other hand, P12 acknowledged having personal blind spots that they are responsible for tackling:``I have blind spots. And my job is, you know, to make sure that I address them and tackle them and make sure that they don't become something that hinders my storytelling ability.'' As a senior researcher with dominant demographic identities, P14 shared how they prioritized being mindful of how their identities are perceived alongside their position and power:  
\begin{quote}
    I worry about being perceived as the White guy coming in to tell you what's right and what's wrong. [...] But I've definitely been in or worked with clients where people or teams, they don't want to work in front of others. And so I try to be conscious of those dynamics and also the fact that just me coming in as an outside person and depending on the group, as a White male coming into a group where I can be perceived as [...] coming into some community.
\end{quote}

Participants shared their practices to center equity in their work. Central to these practices were efforts to build their awareness of personal and data biases and mitigating or being aware of the impact of their positionalities on their design process---which present research opportunities for more engagement between the practitioner and research community around producing equitable visualizations of demographic data. 
\section{Discussion} 
We investigated how the politics, values, beliefs, and positionalities of designers influence their production of visualizations of race and gender, and the challenges they face. Through our seventeen interviews, we found that the production of visualizations of demographic data is deeply impacted by the politics and positionalities of individual designers. We argue that visualizing protected demographic data is a value-laden and political process as a result of interactions between these aforementioned factors. Next, we discuss some implications of our work and the politics implicit in visualization workflows of protected demographic data. Lastly, we highlight the value of centering designers' situated knowledges in addressing the design and data challenges around the creation of public visualization artifacts.

 % Through our seventeen interviews, we found that our participants faced data and design challenges in part due to the sociopolitical nature of demographic categories. Participants also navigated tensions involving the design of accessible and open data visualizations in a climate of misinformation and reflected on their power and politics in their roles as designers. Lastly, our participants shared how their understanding of their identities, alongside their lived experiences and backgrounds, shaped their design approach. Our study revealed that the production of visualizations of demographic data is deeply impacted by the politics, and positionalities of individual designers. We argue that interactions between these factors result in visualizations of demographic data being value-laden, which has important implications for research on representing demographic data about marginalized populations.

% \subsection{Visualizations Are Socially Constructed From the Designer's Perspective}
\subsection{Politics and Situated Knowledges of Practitioners in Visualization Design}
\label{sec:discussion-situated-knowledges}
Our interviews showed how studying data visualizations as separate from their designers neglects important aspects of their process, that are often absent from positivist representations of the visualization design process~\cite{parsons2021understanding, MunznerNested2009}. Particularly, participant responses revealed tensions between their purported role as a neutral communicator of data and their value-laden practices. These tensions reflect Agre's work~\cite{agre1995institutional} on the assumed neutrality of librarians. Similar to librarians, visualization designers are in charge of communicating information to audiences and are implicitly thought to be neutral but have politics and are situated in broader ``institutional circuits'' that influence the artifacts they produce. 

Visualizing demographic data of marginalized populations is deeply political as designers navigate power arrangements in their workplaces and leverage their power and positionalities in their design. We observed how designers with an acute awareness of how they are privileged and/or discriminated against based on their positionalities and identities tended to be more outwardly explicit in their political approaches when visualizing race and gender (\S\ref{sec:politics-power}). Our findings have implications for other forms of demographic data, beyond race and gender, that may have political consequences when visualized. For instance, these could include bar charts of migrant crossings at the US southern border consisting of arbitrary categories that have been circulating to spread misinformation in the 2024 US presidential elections~\cite{TrumpChart-NYT, bump_analysis_2024}, where the immigration-related demographic categories are constructed from arbitrary classification systems. We discuss these implications further in \S\ref{implications:misinfo}.

Importantly, our findings around designer conceptualizations of neutrality, objectivity, and value-laden (\S\ref{sec:tensions-neutrality}) approaches echo the concept of \textbf{situated knowledges}, first proposed by feminist Donna Haraway~\cite{haraway1988situated} when questioning notions of neutrality and objectivity in science. Our participants do not exist in a neutral state but rather have diverse lived experiences and backgrounds that make up their situated knowledges, from which they then approach the production of visualization artifacts. Thus, we offer an empirical study of situated knowledges of visualization designers working with race and gender data to the visualization community. Our work reflects long-established knowledge in communities working with marginalized populations in HCI and beyond and that this study only scratches the surface of how feminist epistemologies can better support equitable visualization research and design practices. We also supplement ~\citet{Liang2022}'s work on the tensions (exploitation, membership, disclosure, and allyship) that HCI researchers encounter when working with marginalized populations. Specifically, our findings focus on the tensions around membership, positionality, and allyship that visualization designers face when representing vulnerable groups. Thus, our work bridges a knowledge gap around the intersecting challenges that both researchers and practitioners within HCI encounter when working with underrepresented populations. 

We also acknowledge that this work is limited in its reproducibility and that our positionalities as researchers with minoritized identities and previous experiences working with protected demographic data also influenced our research questions. Additionally, recruiting designers with minoritized identities was challenging and our sample of participants was skewed towards individuals identifying as White and male. Hence, the findings of this paper should be expanded with a more diverse set of participants. Lastly, by interpreting our findings through the feminist lens of situated knowledges, we identify a series of considerations as to how designer conceptualizations of notions of power, politics, and positionality can support equitable research and design practices on visualizing demographic data, which we discuss in \S\ref{considerations}.

% We began this study with Winner's definition of politics as ``arrangements of power and authority in human associations as well as the activities that take place within those arrangements''~\cite[~p.123]{Winner1980-WINDAH-3}. In this context, we observed that visualizing demographic data of marginalized populations is deeply political as designers navigate power arrangements in their workplaces and leverage their power and positionalities in their design. In particular, our study illustrates how designers with an acute awareness of how they are privileged and/or discriminated against based on their positionalities and identities tended to be more outwardly explicit in their political approaches when visualizing race and gender. Other participants shared how they worked to grow their awareness of how their positionalities influenced their visualization artifacts. % Visualizations are socially constructed as the product of designer choices, which are often influenced by the designer's positionality, values, beliefs, and politics. Participants reflected on their politics, with some questioning the value of neutrality which they viewed as a challenging concept, while others instead chose to prioritize different values. Participants with minoritized identities shared how a neutral approach when visualizing data about marginalized populations could negatively impact these groups. 

\subsection{Challenges of Positivist Approaches to Visualizing Race and Gender}
Our findings suggest that the challenges and harms faced by designers working with race and gender are related to concerns around the social construction of protected demographic categories~\cite{Andrus,Andrus-2022, clair2015sociology, jablonski2021skin}, the lack of support for designers working with these kinds of data~\cite{dhawka2023we, Liang2022, Cabric2024}, and the limitations of positivist approaches that prioritize designer neutrality~\cite{akbaba2024entanglements, meyer-dykes2020}. 

Visualizing race and gender data is uniquely challenging compared to other forms of sociodemographic data~\cite{Cabric2024, dhawka2023we, SchwabishFeng}. Due to the historical nature of demographic classification being used to discriminate against marginalized populations~\cite{clair2015sociology, clark2020disproportionate, jablonski2021skin}, the challenges of visualizing this data may not generalize across countries. For instance, best practices developed in the US may not map to demographic categories elsewhere or may not adequately contextualize distinct histories of discrimination related to protected demographics. Protected demographic categories are also increasingly subject to political decisions such as the addition of the ``MENA'' racial category~\cite{MENA-NYT} and the expansion of gender categories~\cite{censusUSA-Sex} to the 2030 US census. Thus, practitioners often have to account for systemic changes in these categories alongside their histories of discrimination when visually representing them for audiences with varying levels of data and visualization literacies. Another related challenge is the lack of data available for certain racial or gender categories and the unintended side-effects of collecting fine-grained demographic data about marginalized populations. Despite growing interest in collecting more race and gender data~\cite{renzaho2023lack}, we echo the stance of other critical AI and HCI researchers  around critically weighing the harms resulting from further data collection, such as increased surveillance of and algorithmic harm towards marginalized populations~\cite{Andrus-2022, hanna2020towards, Keyes2018}.  

Focusing on equitably visualizing existing data about race and gender, practitioners shared their strategies such as using icons and anthropographics to ``humanize'' the people in the data and the limitations of these approaches. For instance, the creation of anthropographic visualizations is labor intensive and is largely restricted by a lack of tools to rapidly create humanizing representations of data. Similarly, participants shared concerns about how their audiences might interpret complex anthropographic visualizations, such as a lack of consensus on what icons represent and balancing the creation of complex visualizations with audience comprehensibility (\S\ref{sec:design-data-challenges}). Notably, the majority of our participants expressed wanting to design visualizations that their audiences could relate to, while also highlighting their concerns about being an outsider to the groups they are designing for and the being aware of the risks of harmful misrepresentation. Our findings echo the challenges around visualizing race as outlined by~\citet{dhawka2023we}, which present opportunities for visualization researchers to engage in participatory and co-design research with practitioners. Participants were also worried about perpetuating information inequality by designing complex visualizations that would be equally comprehensible to audiences with varying levels of visualization literacy. Here, our findings are relevant for understanding how practitioners account for the perceived data visualization literacies of their audiences and for investigating the challenges of designing public visualizations of race and gender. 

Overall, our findings show how positivist approaches, tools, and workflows for visualizing protected demographic data may be insufficient toward addressing challenges faced by practitioners. Particularly, our participants' understanding of neutrality, objectivity, and other design values underscore the plurality of visualization approaches that diverge from positivist ones that prioritize designer neutrality. Furthermore, our racially minoritized participants' perspectives on the lack of objectivity related to visualizing race and gender resonate with critical data studies work on datafication~\cite{mejias2019datafication} and data colonialism~\cite{couldry2019making} that question the growing power of data, its association with objectivity, and its impact on marginalized groups. Questions on the relationships and tensions around neutrality and objectivity (as described in \S\ref{sec:differences-neutrality-objectivity}), and value-laden approaches (\S\ref{sec:value-laden-approaches}), echo existing concerns in HCI~\cite{bardzell-bardzell-2011}. Engaging with these questions, alongside the notions of  ``situated knowledges''~\cite{haraway1988situated} and ``feminist versions of objectivity''~\cite{haraway_simians_1998} (cited in ~\cite{bardzell-bardzell-2011} via \cite{mcdowell_doing_1997}) in the context of visualization practices can support research on accommodating epistemological pluralism and a critical technical practice of visualization design~\cite{agre2014toward} that prioritize designer values alongside objectivity and reflexivity. 

\section{Considerations for Future Research and Practice}
\label{considerations}
Based on our findings, we propose a set of considerations for future research and practice for the visualization research community to more deeply engage with the experiences of designers working with protected demographic data. 

% \subsection{Safeguarding Visualizations of Protected Demographic Data From Misinformation}
\subsection{Safeguarding Demographic Data Visualizations from Misappropriation and Prioritizing Designer Safety}
\label{implications:misinfo}
% \subsection{Mitigating the Potential for Misleading Visualizations}
Public demographic data visualizations are essential for transparency, disclosure, and reproducibility. However, we observed that our participants were acutely aware of the potential of their work being misappropriated to spread misinformation related to marginalized demographic groups. Several factors make these kinds of demographic data visualizations vulnerable to being misappropriated to spread misinformation about the groups being represented. As prior work by Lee et al.~\cite{lee2021viral} has demonstrated, social media users co-opted mainstream visualization strategies to create counter-data visualizations of open data from US government agencies during the COVID-19 pandemic. Similarly, recent work by Lisnic et al.~\cite{lisnic2023misleading, lisnic2024} illustrated how existing visualizations are often vulnerable to being manipulated toward misleading audiences in misinformation campaigns. Although our findings demonstrate how designers use best practices on reporting census demographic data and protecting the privacy of individuals, these practices often do not include safeguards against misinformation. Thus, we see opportunities for the visualization research community to engage in inter-disciplinary collaborations with ongoing research in the misinformation space to explore tools and strategies to detect and prevent the use of visualizations of protected demographic data in misleading audiences. 

Furthermore, our participants' experiences with online harassment and audience distrust after sharing public visualizations of race and gender underscore the lack of measures to protect them against harassment for creating visualizations of marginalized groups. Here, we see opportunities for research and practice to introduce interventions and strategies that prioritize designer safety and the privacy of the people being visualized while increasing audience trust in data visualizations. Countering the use of visualizations in misinformation campaigns also requires the ability to detect provenance and manipulations made to the original visualizations. Hence, we call for more research exploring new tools and data visualization literacy initiatives to support audiences in identifying and debunking misleading visualizations. 
\subsection{Designing Visualization Tools for Epistemological Pluralism}
Supporting the equitable creation of visualizations of demographic data requires embedding feminist, critical, and alternative epistemologies early in the design process, particularly in visualization tools and pipelines. 
Yet, current tools and software for visualization design often assume a positivist approach that prioritizes design values such as objectivity and neutrality that are likely suited for certain types of data over others. In particular, our participants discussed having to balance designing complex visualizations and retaining an audience's attention while also being aware that standard visualizations (such as barcharts) can be inadequate when communicating demographic data of marginalized populations. Thus, we advocate for the design of visualization tools that consider diverse epistemologies alongside the challenges of representing socially constructed demographic data. As ~\citet{datavis-literacy} argue, an openness to diverse epistemologies, such as feminist theory, can also better support data visualization literacy initiatives by supporting learners in accounting for the social construction of visualizations. Acknowledging that visualizations are socially constructed may allow for mainstream visualization tools to include disclosure features for designers to contextualize issues in the data and foreground the designer's positionality and politics to combat misinformation~\cite{Schwan2022, dork}. Since visualizations can serve multiple purposes, including educational and persuasive ones~\cite{lee-robbins2022}, visualization tools that support diverse critical epistemologies could also provide audiences with multiple views and narratives of the same dataset, including author-driven ones, while also supporting the agency of audiences to explore the data. 

% \section{Acknowledgments}

% re this section as a numbered or unnumbered {\verb|\section|}; please use the ``{\verb|acks|}'' environment.
\section{Conclusion}
In this paper, we presented findings from a qualitative study investigating the challenges and experiences of practitioners visualizing race and gender data. We aimed to explore how the positionalities and politics of designers influence their design process, and how practitioners engage in reflexive, creative strategies to equitably represent protected demographic data. As with any design artifact, data visualizations are socially constructed and reflect the values and choices of their designers. Visualizing data is also an inherently human process of communication. Hence, when representing marginalized populations, the designer's lived experiences and politics, among other factors, influence the end product and have implications for the people being visualized. Our work reveals tensions and open questions that visualization practitioners have to grapple with, and we share our work with the hope that it serves as a starting point for future visualization research that considers how to better support visualization designers working with socially constructed demographic data.

% we can better support designers in equitably representing protected demographic data.  
% \section{Appendices}

\begin{acks}
We are grateful to our participants for engaging deeply with us during our interviews and to Joice Tang, Samuel So, Anoolia Gakhokidze, McKane Andrus, and members of the Learning, Epistemology, and Design Lab for their advice and feedback on drafts of this paper. The authors also thank our anonymous reviewers for their thoughtful comments and suggestions.
\end{acks}

\bibliographystyle{ACM-Reference-Format}
\bibliography{references}


\begin{thebibliography}{71}

%%% the Web address.

\ifx \showCODEN    \undefined \def \showCODEN     #1{\unskip}     \fi
\ifx \showDOI      \undefined \def \showDOI       #1{#1}\fi
\ifx \showISBNx    \undefined \def \showISBNx     #1{\unskip}     \fi
\ifx \showISBNxiii \undefined \def \showISBNxiii  #1{\unskip}     \fi
\ifx \showISSN     \undefined \def \showISSN      #1{\unskip}     \fi
\ifx \showLCCN     \undefined \def \showLCCN      #1{\unskip}     \fi
\ifx \shownote     \undefined \def \shownote      #1{#1}          \fi
\ifx \showarticletitle \undefined \def \showarticletitle #1{#1}   \fi
\ifx \showURL      \undefined \def \showURL       {\relax}        \fi
% invisible to TeX
\providecommand\bibfield[2]{#2}
\providecommand\bibinfo[2]{#2}
\providecommand\natexlab[1]{#1}
\providecommand\showeprint[2][]{arXiv:#2}

\bibitem[Agre(1995)]%
        {agre1995institutional}
\bibfield{author}{\bibinfo{person}{Philip~E Agre}.} \bibinfo{year}{1995}\natexlab{}.
\newblock \showarticletitle{Institutional circuitry: Thinking about the forms and uses of information}.
\newblock \bibinfo{journal}{\emph{Information Technology and Libraries}} \bibinfo{volume}{14}, \bibinfo{number}{4} (\bibinfo{year}{1995}), \bibinfo{pages}{225--230}.
\newblock


\bibitem[Agre(2014)]%
        {agre2014toward}
\bibfield{author}{\bibinfo{person}{Philip~E Agre}.} \bibinfo{year}{2014}\natexlab{}.
\newblock \showarticletitle{Toward a critical technical practice: Lessons learned in trying to reform AI}.
\newblock In \bibinfo{booktitle}{\emph{Social science, technical systems, and cooperative work}}. \bibinfo{publisher}{Psychology Press}, \bibinfo{pages}{131--157}.
\newblock


\bibitem[Akbaba et~al\mbox{.}(2024)]%
        {akbaba2024entanglements}
\bibfield{author}{\bibinfo{person}{Derya Akbaba}, \bibinfo{person}{Lauren Klein}, {and} \bibinfo{person}{Miriah Meyer}.} \bibinfo{year}{2024}\natexlab{}.
\newblock \showarticletitle{Entanglements for Visualization: Changing Research Outcomes through Feminist Theory}.
\newblock \bibinfo{journal}{\emph{IEEE Transactions on Visualization and Computer Graphics}} \bibinfo{volume}{31}, \bibinfo{number}{1} (\bibinfo{year}{2024}), \bibinfo{numpages}{11}~pages.
\newblock
\urldef\tempurl%
\url{https://doi.org/10.1109/TVCG.2024.3456171}
\showDOI{\tempurl}


\bibitem[Akbaba et~al\mbox{.}(2023)]%
        {akbaba2023troubling}
\bibfield{author}{\bibinfo{person}{Derya Akbaba}, \bibinfo{person}{Devin Lange}, \bibinfo{person}{Michael Correll}, \bibinfo{person}{Alexander Lex}, {and} \bibinfo{person}{Miriah Meyer}.} \bibinfo{year}{2023}\natexlab{}.
\newblock \showarticletitle{Troubling collaboration: Matters of care for visualization design study}. In \bibinfo{booktitle}{\emph{Proceedings of the 2023 CHI Conference on Human Factors in Computing Systems}} (Hamburg, Germany) \emph{(\bibinfo{series}{CHI '23})}. \bibinfo{publisher}{Association for Computing Machinery}, \bibinfo{address}{New York, NY, USA}, \bibinfo{numpages}{14}~pages.
\newblock
\urldef\tempurl%
\url{https://doi.org/10.1145/3544548.3581168}
\showDOI{\tempurl}


\bibitem[Andrus et~al\mbox{.}(2021)]%
        {Andrus}
\bibfield{author}{\bibinfo{person}{McKane Andrus}, \bibinfo{person}{Elena Spitzer}, \bibinfo{person}{Jeffrey Brown}, {and} \bibinfo{person}{Alice Xiang}.} \bibinfo{year}{2021}\natexlab{}.
\newblock \showarticletitle{What We Can't Measure, We Can't Understand: Challenges to Demographic Data Procurement in the Pursuit of Fairness}. In \bibinfo{booktitle}{\emph{Proceedings of the 2021 ACM Conference on Fairness, Accountability, and Transparency}} (Virtual Event, Canada) \emph{(\bibinfo{series}{FAccT '21})}. \bibinfo{publisher}{Association for Computing Machinery}, \bibinfo{address}{New York, NY, USA}, \bibinfo{pages}{249–260}.
\newblock
\showISBNx{9781450383097}
\urldef\tempurl%
\url{https://doi.org/10.1145/3442188.3445888}
\showDOI{\tempurl}


\bibitem[Andrus and Villeneuve(2022)]%
        {Andrus-2022}
\bibfield{author}{\bibinfo{person}{McKane Andrus} {and} \bibinfo{person}{Sarah Villeneuve}.} \bibinfo{year}{2022}\natexlab{}.
\newblock \showarticletitle{Demographic-Reliant Algorithmic Fairness: Characterizing the Risks of Demographic Data Collection in the Pursuit of Fairness}. In \bibinfo{booktitle}{\emph{Proceedings of the 2022 ACM Conference on Fairness, Accountability, and Transparency}} (Seoul, Republic of Korea) \emph{(\bibinfo{series}{FAccT '22})}. \bibinfo{publisher}{Association for Computing Machinery}, \bibinfo{address}{New York, NY, USA}, \bibinfo{pages}{1709–1721}.
\newblock
\showISBNx{9781450393522}
\urldef\tempurl%
\url{https://doi.org/10.1145/3531146.3533226}
\showDOI{\tempurl}


\bibitem[Baah et~al\mbox{.}(2019)]%
        {baah2019marginalization}
\bibfield{author}{\bibinfo{person}{Foster~Osei Baah}, \bibinfo{person}{Anne~M Teitelman}, {and} \bibinfo{person}{Barbara Riegel}.} \bibinfo{year}{2019}\natexlab{}.
\newblock \showarticletitle{{Marginalization: Conceptualizing Patient Vulnerabilities in the Framework of Social Determinants of Health—An integrative Review}}.
\newblock \bibinfo{journal}{\emph{{Nursing Inquiry}}} \bibinfo{volume}{26}, \bibinfo{number}{1} (\bibinfo{year}{2019}), \bibinfo{pages}{e12268}.
\newblock
\urldef\tempurl%
\url{https://doi.org/10.1111/nin.12268}
\showDOI{\tempurl}


\bibitem[Bako et~al\mbox{.}(2023)]%
        {bako2022understanding}
\bibfield{author}{\bibinfo{person}{Hannah~K. Bako}, \bibinfo{person}{Xinyi Liu}, \bibinfo{person}{Leilani Battle}, {and} \bibinfo{person}{Zhicheng Liu}.} \bibinfo{year}{2023}\natexlab{}.
\newblock \showarticletitle{Understanding how Designers Find and Use Data Visualization Examples}.
\newblock \bibinfo{journal}{\emph{IEEE Transactions on Visualization and Computer Graphics}} \bibinfo{volume}{29}, \bibinfo{number}{1} (\bibinfo{year}{2023}), \bibinfo{pages}{1048--1058}.
\newblock
\urldef\tempurl%
\url{https://doi.org/10.1109/TVCG.2022.3209490}
\showDOI{\tempurl}


\bibitem[Bardzell and Bardzell(2011)]%
        {bardzell-bardzell-2011}
\bibfield{author}{\bibinfo{person}{Shaowen Bardzell} {and} \bibinfo{person}{Jeffrey Bardzell}.} \bibinfo{year}{2011}\natexlab{}.
\newblock \showarticletitle{Towards a feminist HCI methodology: social science, feminism, and HCI}. In \bibinfo{booktitle}{\emph{Proceedings of the SIGCHI Conference on Human Factors in Computing Systems}} (Vancouver, BC, Canada) \emph{(\bibinfo{series}{CHI '11})}. \bibinfo{publisher}{Association for Computing Machinery}, \bibinfo{address}{New York, NY, USA}, \bibinfo{pages}{675–684}.
\newblock
\showISBNx{9781450302289}
\urldef\tempurl%
\url{https://doi.org/10.1145/1978942.1979041}
\showDOI{\tempurl}


\bibitem[Berret and Munzner(2024)]%
        {Berret-Munzner2024}
\bibfield{author}{\bibinfo{person}{Charles Berret} {and} \bibinfo{person}{Tamara Munzner}.} \bibinfo{year}{2024}\natexlab{}.
\newblock \showarticletitle{Iceberg Sensemaking: A Process Model for Critical Data Analysis}.
\newblock \bibinfo{journal}{\emph{IEEE Transactions on Visualization and Computer Graphics}} (\bibinfo{year}{2024}), \bibinfo{pages}{1--18}.
\newblock
\urldef\tempurl%
\url{https://doi.org/10.1109/TVCG.2024.3486613}
\showDOI{\tempurl}


\bibitem[Boy et~al\mbox{.}(2017)]%
        {Boy}
\bibfield{author}{\bibinfo{person}{Jeremy Boy}, \bibinfo{person}{Anshul~Vikram Pandey}, \bibinfo{person}{John Emerson}, \bibinfo{person}{Margaret Satterthwaite}, \bibinfo{person}{Oded Nov}, {and} \bibinfo{person}{Enrico Bertini}.} \bibinfo{year}{2017}\natexlab{}.
\newblock \showarticletitle{{Showing People Behind Data: Does Anthropomorphizing Visualizations Elicit More Empathy for Human Rights Data?}}. In \bibinfo{booktitle}{\emph{Proceedings of the 2017 CHI Conference on Human Factors in Computing Systems}} (Denver, Colorado, USA) \emph{(\bibinfo{series}{CHI '17})}. \bibinfo{publisher}{Association for Computing Machinery}, \bibinfo{address}{New York, NY, USA}, \bibinfo{pages}{5462–5474}.
\newblock
\showISBNx{9781450346559}
\urldef\tempurl%
\url{https://doi.org/10.1145/3025453.3025512}
\showDOI{\tempurl}


\bibitem[Bump(2024)]%
        {bump_analysis_2024}
\bibfield{author}{\bibinfo{person}{Philip Bump}.} \bibinfo{year}{2024}\natexlab{}.
\newblock \showarticletitle{Analysis {\textbar} {A} look at {Trump}’s misleading, inaccurate graph of {U}.{S}. immigration}.
\newblock \bibinfo{journal}{\emph{The Washington Post}} (\bibinfo{date}{May} \bibinfo{year}{2024}).
\newblock
\showISSN{0190-8286}
\urldef\tempurl%
\url{https://www.washingtonpost.com/politics/2024/05/23/look-trumps-misleading-inaccurate-graph-us-immigration/}
\showURL{%
\tempurl}


\bibitem[Cabric et~al\mbox{.}(2023)]%
        {cabric2023open}
\bibfield{author}{\bibinfo{person}{Florent Cabric}, \bibinfo{person}{Margr{\'e}t~Vilborg Bjarnad{\'o}ttir}, \bibinfo{person}{Anne-Flore Cabouat}, {and} \bibinfo{person}{Petra Isenberg}.} \bibinfo{year}{2023}\natexlab{}.
\newblock \showarticletitle{Open Questions about the Visualization of Sociodemographic Data}. In \bibinfo{booktitle}{\emph{2023 IEEE Workshop on Visualization for Social Good (VIS4Good)}}. IEEE, \bibinfo{pages}{16--20}.
\newblock
\urldef\tempurl%
\url{https://doi.org/10.1109/VIS4Good60218.2023.00010}
\showDOI{\tempurl}


\bibitem[Cabric et~al\mbox{.}(2024)]%
        {Cabric2024}
\bibfield{author}{\bibinfo{person}{Florent Cabric}, \bibinfo{person}{Margrét~Vilborg Bjarnadóttir}, \bibinfo{person}{Meng Ling}, \bibinfo{person}{Guðbjörg~Linda Rafnsdóttir}, {and} \bibinfo{person}{Petra Isenberg}.} \bibinfo{year}{2024}\natexlab{}.
\newblock \showarticletitle{Eleven Years of Gender Data Visualization: A Step Towards More Inclusive Gender Representation}.
\newblock \bibinfo{journal}{\emph{IEEE Transactions on Visualization and Computer Graphics}} \bibinfo{volume}{30}, \bibinfo{number}{1} (\bibinfo{year}{2024}), \bibinfo{pages}{316--326}.
\newblock
\urldef\tempurl%
\url{https://doi.org/10.1109/TVCG.2023.3327369}
\showDOI{\tempurl}


\bibitem[{Census Bureau}(1997)]%
        {censusUSA}
\bibfield{author}{\bibinfo{person}{{Census Bureau}}.} \bibinfo{year}{1997}\natexlab{}.
\newblock \bibinfo{title}{{About the Topic of Race}}.
\newblock
\newblock
\urldef\tempurl%
\url{https://perma.cc/WH3H-X76F}
\showURL{%
Retrieved September 14, 2023 from \tempurl}


\bibitem[{Census Bureau}(2021)]%
        {censusUSA-Sex}
\bibfield{author}{\bibinfo{person}{{Census Bureau}}.} \bibinfo{year}{2021}\natexlab{}.
\newblock \bibinfo{title}{{About Age and Sex}}.
\newblock
\newblock
\urldef\tempurl%
\url{https://www.census.gov/topics/population/age-and-sex/about.html}
\showURL{%
Retrieved December 10, 2024 from \tempurl}


\bibitem[Chen et~al\mbox{.}(2023)]%
        {Chen2023}
\bibfield{author}{\bibinfo{person}{Yiqun~T. Chen}, \bibinfo{person}{Angela D.~R. Smith}, \bibinfo{person}{Katharina Reinecke}, {and} \bibinfo{person}{Alexandra To}.} \bibinfo{year}{2023}\natexlab{}.
\newblock \showarticletitle{Why, when, and from whom: considerations for collecting and reporting race and ethnicity data in HCI}. In \bibinfo{booktitle}{\emph{Proceedings of the 2023 CHI Conference on Human Factors in Computing Systems}} (Hamburg, Germany) \emph{(\bibinfo{series}{CHI '23})}. \bibinfo{publisher}{Association for Computing Machinery}, \bibinfo{address}{New York, NY, USA}, Article \bibinfo{articleno}{395}, \bibinfo{numpages}{15}~pages.
\newblock
\showISBNx{9781450394215}
\urldef\tempurl%
\url{https://doi.org/10.1145/3544548.3581122}
\showDOI{\tempurl}


\bibitem[Clair and Denis(2015)]%
        {clair2015sociology}
\bibfield{author}{\bibinfo{person}{Matthew Clair} {and} \bibinfo{person}{Jeffrey~S Denis}.} \bibinfo{year}{2015}\natexlab{}.
\newblock \showarticletitle{{Sociology of Racism}}.
\newblock \bibinfo{journal}{\emph{{The International Encyclopedia of the Social and Behavioral Sciences}}}  \bibinfo{volume}{19} (\bibinfo{year}{2015}), \bibinfo{pages}{857--63}.
\newblock
\urldef\tempurl%
\url{https://projects.iq.harvard.edu/files/deib-explorer/files/sociology_of_racism.pdf}
\showURL{%
\tempurl}


\bibitem[Clark et~al\mbox{.}(2020)]%
        {clark2020disproportionate}
\bibfield{author}{\bibinfo{person}{Eva Clark}, \bibinfo{person}{Karla Fredricks}, \bibinfo{person}{Laila Woc-Colburn}, \bibinfo{person}{Maria~Elena Bottazzi}, {and} \bibinfo{person}{Jill Weatherhead}.} \bibinfo{year}{2020}\natexlab{}.
\newblock \showarticletitle{{Disproportionate Impact of the COVID-19 Pandemic on Immigrant Communities in the United States}}.
\newblock \bibinfo{journal}{\emph{PLoS Neglected Tropical Diseases}} \bibinfo{volume}{14}, \bibinfo{number}{7} (\bibinfo{year}{2020}), \bibinfo{pages}{e0008484}.
\newblock
\urldef\tempurl%
\url{https://doi.org/10.1371/journal.pntd.0008484}
\showURL{%
\tempurl}


\bibitem[Clarke and Braun(2017)]%
        {clarke2017thematic}
\bibfield{author}{\bibinfo{person}{Victoria Clarke} {and} \bibinfo{person}{Virginia Braun}.} \bibinfo{year}{2017}\natexlab{}.
\newblock \showarticletitle{Thematic analysis}.
\newblock \bibinfo{journal}{\emph{The journal of positive psychology}} \bibinfo{volume}{12}, \bibinfo{number}{3} (\bibinfo{year}{2017}), \bibinfo{pages}{297--298}.
\newblock


\bibitem[Commission(1966)]%
        {USEEOC}
\bibfield{author}{\bibinfo{person}{United States Equal Employment~Opportunity Commission}.} \bibinfo{year}{1966}\natexlab{}.
\newblock \bibinfo{title}{{EEO-1 Component 1 Data Collection}}.
\newblock
\newblock
\urldef\tempurl%
\url{https://perma.cc/388J-JZV5}
\showURL{%
Retrieved September 9, 2024 from \tempurl}


\bibitem[Correll(2019)]%
        {Correll}
\bibfield{author}{\bibinfo{person}{Michael Correll}.} \bibinfo{year}{2019}\natexlab{}.
\newblock \showarticletitle{{Ethical Dimensions of Visualization Research}}. In \bibinfo{booktitle}{\emph{Proceedings of the 2019 CHI Conference on Human Factors in Computing Systems}} (Glasgow, Scotland Uk) \emph{(\bibinfo{series}{CHI '19})}. \bibinfo{publisher}{Association for Computing Machinery}, \bibinfo{address}{New York, NY, USA}, \bibinfo{pages}{1–13}.
\newblock
\showISBNx{9781450359702}
\urldef\tempurl%
\url{https://doi.org/10.1145/3290605.3300418}
\showDOI{\tempurl}


\bibitem[Couldry and Mejias(2019)]%
        {couldry2019making}
\bibfield{author}{\bibinfo{person}{Nick Couldry} {and} \bibinfo{person}{Ulises Mejias}.} \bibinfo{year}{2019}\natexlab{}.
\newblock \showarticletitle{Making data colonialism liveable: how might data’s social order be regulated?}
\newblock \bibinfo{journal}{\emph{Internet Policy Review}} \bibinfo{volume}{8}, \bibinfo{number}{2} (\bibinfo{year}{2019}), \bibinfo{pages}{1--16}.
\newblock


\bibitem[Dhawka et~al\mbox{.}(2022)]%
        {dhawka2022representing}
\bibfield{author}{\bibinfo{person}{Priya Dhawka}, \bibinfo{person}{Helen~Ai He}, {and} \bibinfo{person}{Wesley Willett}.} \bibinfo{year}{2022}\natexlab{}.
\newblock \showarticletitle{Representing marginalized populations: Challenges in anthropographics}.
\newblock \bibinfo{journal}{\emph{arXiv preprint arXiv:2210.02660}} (\bibinfo{year}{2022}).
\newblock


\bibitem[Dhawka et~al\mbox{.}(2023)]%
        {dhawka2023we}
\bibfield{author}{\bibinfo{person}{Priya Dhawka}, \bibinfo{person}{Helen~Ai He}, {and} \bibinfo{person}{Wesley Willett}.} \bibinfo{year}{2023}\natexlab{}.
\newblock \showarticletitle{We Are the Data: Challenges and Opportunities for Creating Demographically Diverse Anthropographics}. In \bibinfo{booktitle}{\emph{Proceedings of the 2023 CHI Conference on Human Factors in Computing Systems}} (Hamburg, Germany) \emph{(\bibinfo{series}{CHI '23})}. \bibinfo{publisher}{Association for Computing Machinery}, \bibinfo{address}{New York, NY, USA}, Article \bibinfo{articleno}{807}, \bibinfo{numpages}{14}~pages.
\newblock
\showISBNx{9781450394215}
\urldef\tempurl%
\url{https://doi.org/10.1145/3544548.3581086}
\showDOI{\tempurl}


\bibitem[Dhawka et~al\mbox{.}(2024)]%
        {dhawka2024}
\bibfield{author}{\bibinfo{person}{Priya Dhawka}, \bibinfo{person}{Lauren Perera}, {and} \bibinfo{person}{Wesley Willett}.} \bibinfo{year}{2024}\natexlab{}.
\newblock \showarticletitle{Better Little People Pictures: Generative Creation of Demographically Diverse Anthropographics}. In \bibinfo{booktitle}{\emph{Proceedings of the CHI Conference on Human Factors in Computing Systems}} (Honolulu, HI, USA) \emph{(\bibinfo{series}{CHI '24})}. \bibinfo{publisher}{Association for Computing Machinery}, \bibinfo{address}{New York, NY, USA}, Article \bibinfo{articleno}{557}, \bibinfo{numpages}{14}~pages.
\newblock
\showISBNx{9798400703300}
\urldef\tempurl%
\url{https://doi.org/10.1145/3613904.3641957}
\showDOI{\tempurl}


\bibitem[D'Ignazio and Klein(2020)]%
        {datafem}
\bibfield{author}{\bibinfo{person}{Catherine D'Ignazio} {and} \bibinfo{person}{Lauren~F Klein}.} \bibinfo{year}{{2020}}\natexlab{}.
\newblock \showarticletitle{{1. The Power Chapter}}.
\newblock In \bibinfo{booktitle}{\emph{{Data Feminism}}}. \bibinfo{publisher}{{The MIT Press}}, \bibinfo{address}{Cambridge, Massachusetts}.
\newblock


\bibitem[D\"{o}rk et~al\mbox{.}(2013)]%
        {dork}
\bibfield{author}{\bibinfo{person}{Marian D\"{o}rk}, \bibinfo{person}{Patrick Feng}, \bibinfo{person}{Christopher Collins}, {and} \bibinfo{person}{Sheelagh Carpendale}.} \bibinfo{year}{2013}\natexlab{}.
\newblock \showarticletitle{{Critical InfoVis: Exploring the Politics of Visualization}}. In \bibinfo{booktitle}{\emph{CHI '13 Extended Abstracts on Human Factors in Computing Systems}} (Paris, France) \emph{(\bibinfo{series}{CHI EA '13})}. \bibinfo{publisher}{Association for Computing Machinery}, \bibinfo{address}{New York, NY, USA}, \bibinfo{pages}{2189–2198}.
\newblock
\showISBNx{9781450319522}
\urldef\tempurl%
\url{https://doi.org/10.1145/2468356.2468739}
\showDOI{\tempurl}


\bibitem[Duarte and Baranauskas(2016)]%
        {duarte2016}
\bibfield{author}{\bibinfo{person}{Emanuel~Felipe Duarte} {and} \bibinfo{person}{M.~Cec\'{\i}lia~C. Baranauskas}.} \bibinfo{year}{2016}\natexlab{}.
\newblock \showarticletitle{Revisiting the Three HCI Waves: A Preliminary Discussion on Philosophy of Science and Research Paradigms}. In \bibinfo{booktitle}{\emph{Proceedings of the 15th Brazilian Symposium on Human Factors in Computing Systems}} (S\~{a}o Paulo, Brazil) \emph{(\bibinfo{series}{IHC '16})}. \bibinfo{publisher}{Association for Computing Machinery}, \bibinfo{address}{New York, NY, USA}, Article \bibinfo{articleno}{38}, \bibinfo{numpages}{4}~pages.
\newblock
\showISBNx{9781450352352}
\urldef\tempurl%
\url{https://doi.org/10.1145/3033701.3033740}
\showDOI{\tempurl}


\bibitem[D’Ignazio and Bhargava(2020)]%
        {datavis-literacy}
\bibfield{author}{\bibinfo{person}{Catherine D’Ignazio} {and} \bibinfo{person}{Rahul Bhargava}.} \bibinfo{year}{2020}\natexlab{}.
\newblock \showarticletitle{13. {Data} visualization literacy: {A} feminist starting point}.
\newblock In \bibinfo{booktitle}{\emph{Data {Visualization} in {Society}}}, \bibfield{editor}{\bibinfo{person}{Martin Engebretsen} {and} \bibinfo{person}{Helen Kennedy}} (Eds.). \bibinfo{publisher}{Amsterdam University Press}, \bibinfo{pages}{207--222}.
\newblock
\showISBNx{978-90-485-4313-7}
\urldef\tempurl%
\url{https://doi.org/10.1515/9789048543137-017}
\showDOI{\tempurl}


\bibitem[Elli et~al\mbox{.}(2022)]%
        {elli2022visualizing}
\bibfield{author}{\bibinfo{person}{Tommaso Elli}, \bibinfo{person}{Adam Bradley}, \bibinfo{person}{Uta Hinrichs}, {and} \bibinfo{person}{Christopher Collins}.} \bibinfo{year}{2022}\natexlab{}.
\newblock \showarticletitle{Visualizing stories of sexual harassment in the academy: {Community} empowerment through qualitative data}.
\newblock \bibinfo{journal}{\emph{DRS Biennial Conference Series}} (\bibinfo{date}{June} \bibinfo{year}{2022}).
\newblock
\urldef\tempurl%
\url{https://dl.designresearchsociety.org/drs-conference-papers/drs2022/researchpapers/134}
\showURL{%
\tempurl}


\bibitem[Feinberg(2022)]%
        {feinberg2022everyday}
\bibfield{author}{\bibinfo{person}{Melanie Feinberg}.} \bibinfo{year}{2022}\natexlab{}.
\newblock \bibinfo{booktitle}{\emph{Everyday Adventures with Unruly Data}}.
\newblock \bibinfo{publisher}{MIT Press}, \bibinfo{address}{Cambridge, Massachusetts}.
\newblock


\bibitem[Fichera(2024)]%
        {TrumpChart-NYT}
\bibfield{author}{\bibinfo{person}{Angelo Fichera}.} \bibinfo{year}{2024}\natexlab{}.
\newblock \bibinfo{title}{{Fact-Checking the Immigration Chart That Trump Says ‘Saved My Life’}}.
\newblock \bibinfo{howpublished}{Available: \url{https://www.nytimes.com/2024/07/19/us/politics/trump-immigration-chart-fact-check.html}}.
\newblock


\bibitem[Forest(2005)]%
        {forest2005changing}
\bibfield{author}{\bibinfo{person}{Benjamin Forest}.} \bibinfo{year}{2005}\natexlab{}.
\newblock \showarticletitle{The changing demographic, legal, and technological contexts of political representation}.
\newblock \bibinfo{journal}{\emph{Proceedings of the National Academy of Sciences}} \bibinfo{volume}{102}, \bibinfo{number}{43} (\bibinfo{year}{2005}), \bibinfo{pages}{15331--15336}.
\newblock


\bibitem[Hanna et~al\mbox{.}(2020)]%
        {hanna2020towards}
\bibfield{author}{\bibinfo{person}{Alex Hanna}, \bibinfo{person}{Emily Denton}, \bibinfo{person}{Andrew Smart}, {and} \bibinfo{person}{Jamila Smith-Loud}.} \bibinfo{year}{2020}\natexlab{}.
\newblock \showarticletitle{Towards a critical race methodology in algorithmic fairness}. In \bibinfo{booktitle}{\emph{Proceedings of the 2020 conference on fairness, accountability, and transparency}}. \bibinfo{publisher}{Association for Computing Machinery}, \bibinfo{address}{New York, NY, USA}, \bibinfo{pages}{501--512}.
\newblock


\bibitem[Haraway(1988)]%
        {haraway1988situated}
\bibfield{author}{\bibinfo{person}{Donna Haraway}.} \bibinfo{year}{1988}\natexlab{}.
\newblock \showarticletitle{Situated Knowledges: The Science Question in Feminism and the Privilege of Partial Perspective}.
\newblock \bibinfo{journal}{\emph{Feminist Studies}} \bibinfo{volume}{14}, \bibinfo{number}{3} (\bibinfo{year}{1988}), \bibinfo{pages}{575--599}.
\newblock
\showISSN{00463663}
\urldef\tempurl%
\url{http://www.jstor.org/stable/3178066}
\showURL{%
\tempurl}


\bibitem[Haraway(1998)]%
        {haraway_simians_1998}
\bibfield{author}{\bibinfo{person}{Donna Haraway}.} \bibinfo{year}{1998}\natexlab{}.
\newblock \bibinfo{booktitle}{\emph{Simians, cyborgs, and women: the reinvention of nature} (\bibinfo{edition}{reprinted} ed.)}.
\newblock \bibinfo{publisher}{FAB, Free Association Books}, \bibinfo{address}{London}.
\newblock
\showISBNx{978-1-85343-139-5}


\bibitem[He et~al\mbox{.}(2024)]%
        {he2024}
\bibfield{author}{\bibinfo{person}{Helen~Ai He}, \bibinfo{person}{Jagoda Walny}, \bibinfo{person}{Sonja Thoma}, \bibinfo{person}{Sheelagh Carpendale}, {and} \bibinfo{person}{Wesley Willett}.} \bibinfo{year}{2024}\natexlab{}.
\newblock \showarticletitle{Enthusiastic and Grounded, Avoidant and Cautious: Understanding Public Receptivity to Data and Visualizations}.
\newblock \bibinfo{journal}{\emph{IEEE Transactions on Visualization and Computer Graphics}} \bibinfo{volume}{30}, \bibinfo{number}{1} (\bibinfo{year}{2024}), \bibinfo{pages}{1435--1445}.
\newblock
\urldef\tempurl%
\url{https://doi.org/10.1109/TVCG.2023.3326917}
\showDOI{\tempurl}


\bibitem[Jablonski(2021)]%
        {jablonski2021skin}
\bibfield{author}{\bibinfo{person}{Nina~G Jablonski}.} \bibinfo{year}{2021}\natexlab{}.
\newblock \showarticletitle{{Skin Color and Race}}.
\newblock \bibinfo{journal}{\emph{{American Journal of Physical Anthropology}}} \bibinfo{volume}{175}, \bibinfo{number}{2} (\bibinfo{year}{2021}), \bibinfo{pages}{437--447}.
\newblock
\urldef\tempurl%
\url{https://doi.org/10.1002/ajpa.24200}
\showURL{%
\tempurl}


\bibitem[Keyes(2018)]%
        {Keyes2018}
\bibfield{author}{\bibinfo{person}{Os Keyes}.} \bibinfo{year}{2018}\natexlab{}.
\newblock \showarticletitle{The Misgendering Machines: Trans/HCI Implications of Automatic Gender Recognition}.
\newblock \bibinfo{journal}{\emph{Proc. ACM Hum.-Comput. Interact.}} \bibinfo{volume}{2}, \bibinfo{number}{CSCW}, Article \bibinfo{articleno}{88} (\bibinfo{date}{Nov.} \bibinfo{year}{2018}), \bibinfo{numpages}{22}~pages.
\newblock
\urldef\tempurl%
\url{https://doi.org/10.1145/3274357}
\showDOI{\tempurl}


\bibitem[Lee et~al\mbox{.}(2021)]%
        {lee2021viral}
\bibfield{author}{\bibinfo{person}{Crystal Lee}, \bibinfo{person}{Tanya Yang}, \bibinfo{person}{Gabrielle~D Inchoco}, \bibinfo{person}{Graham~M. Jones}, {and} \bibinfo{person}{Arvind Satyanarayan}.} \bibinfo{year}{2021}\natexlab{}.
\newblock \showarticletitle{Viral Visualizations: How Coronavirus Skeptics Use Orthodox Data Practices to Promote Unorthodox Science Online}. In \bibinfo{booktitle}{\emph{Proceedings of the 2021 CHI Conference on Human Factors in Computing Systems}} (Yokohama, Japan) \emph{(\bibinfo{series}{CHI '21})}. \bibinfo{publisher}{Association for Computing Machinery}, \bibinfo{address}{New York, NY, USA}, Article \bibinfo{articleno}{607}, \bibinfo{numpages}{18}~pages.
\newblock
\showISBNx{9781450380966}
\urldef\tempurl%
\url{https://doi.org/10.1145/3411764.3445211}
\showDOI{\tempurl}


\bibitem[Lee-Robbins and Adar(2022)]%
        {lee-robbins2022}
\bibfield{author}{\bibinfo{person}{Elsie Lee-Robbins} {and} \bibinfo{person}{Eytan Adar}.} \bibinfo{year}{2022}\natexlab{}.
\newblock \showarticletitle{{Affective Learning Objectives for Communicative Visualizations}}.
\newblock \bibinfo{journal}{\emph{IEEE Transactions on Visualization and Computer Graphics}} \bibinfo{volume}{29}, \bibinfo{number}{1} (\bibinfo{year}{2022}), \bibinfo{pages}{1--11}.
\newblock
\urldef\tempurl%
\url{https://doi.org/10.1109/TVCG.2022.3209500}
\showDOI{\tempurl}


\bibitem[Liang et~al\mbox{.}(2021)]%
        {Liang2022}
\bibfield{author}{\bibinfo{person}{Calvin~A. Liang}, \bibinfo{person}{Sean~A. Munson}, {and} \bibinfo{person}{Julie~A. Kientz}.} \bibinfo{year}{2021}\natexlab{}.
\newblock \showarticletitle{{Embracing Four Tensions in Human-Computer Interaction Research with Marginalized People}}.
\newblock \bibinfo{journal}{\emph{ACM Transactions on Computer-Human Interaction (TOCHI)}} \bibinfo{volume}{28}, \bibinfo{number}{2}, Article \bibinfo{articleno}{14} (\bibinfo{year}{2021}), \bibinfo{numpages}{47}~pages.
\newblock
\showISSN{1073-0516}
\urldef\tempurl%
\url{https://doi.org/10.1145/3443686}
\showDOI{\tempurl}


\bibitem[Linxen et~al\mbox{.}(2021)]%
        {WEIRDCHI}
\bibfield{author}{\bibinfo{person}{Sebastian Linxen}, \bibinfo{person}{Christian Sturm}, \bibinfo{person}{Florian Br\"{u}hlmann}, \bibinfo{person}{Vincent Cassau}, \bibinfo{person}{Klaus Opwis}, {and} \bibinfo{person}{Katharina Reinecke}.} \bibinfo{year}{2021}\natexlab{}.
\newblock \showarticletitle{{How WEIRD is CHI?}}. In \bibinfo{booktitle}{\emph{Proceedings of the 2021 CHI Conference on Human Factors in Computing Systems}} (Yokohama, Japan) \emph{(\bibinfo{series}{CHI '21})}. \bibinfo{publisher}{Association for Computing Machinery}, \bibinfo{address}{New York, NY, USA}, Article \bibinfo{articleno}{143}, \bibinfo{numpages}{14}~pages.
\newblock
\urldef\tempurl%
\url{https://doi.org/10.1145/3411764.3445488}
\showDOI{\tempurl}


\bibitem[Lisnic et~al\mbox{.}(2024)]%
        {lisnic2024}
\bibfield{author}{\bibinfo{person}{Maxim Lisnic}, \bibinfo{person}{Alexander Lex}, {and} \bibinfo{person}{Marina Kogan}.} \bibinfo{year}{2024}\natexlab{}.
\newblock \showarticletitle{"Yeah, this graph doesn't show that": Analysis of Online Engagement with Misleading Data Visualizations}. In \bibinfo{booktitle}{\emph{Proceedings of the CHI Conference on Human Factors in Computing Systems}} (Honolulu, HI, USA) \emph{(\bibinfo{series}{CHI '24})}. \bibinfo{publisher}{Association for Computing Machinery}, \bibinfo{address}{New York, NY, USA}, Article \bibinfo{articleno}{199}, \bibinfo{numpages}{14}~pages.
\newblock
\showISBNx{9798400703300}
\urldef\tempurl%
\url{https://doi.org/10.1145/3613904.3642448}
\showDOI{\tempurl}


\bibitem[Lisnic et~al\mbox{.}(2023)]%
        {lisnic2023misleading}
\bibfield{author}{\bibinfo{person}{Maxim Lisnic}, \bibinfo{person}{Cole Polychronis}, \bibinfo{person}{Alexander Lex}, {and} \bibinfo{person}{Marina Kogan}.} \bibinfo{year}{2023}\natexlab{}.
\newblock \showarticletitle{Misleading Beyond Visual Tricks: How People Actually Lie with Charts}. In \bibinfo{booktitle}{\emph{Proceedings of the 2023 CHI Conference on Human Factors in Computing Systems}} (Hamburg, Germany) \emph{(\bibinfo{series}{CHI '23})}. \bibinfo{publisher}{Association for Computing Machinery}, \bibinfo{address}{New York, NY, USA}, Article \bibinfo{articleno}{817}, \bibinfo{numpages}{21}~pages.
\newblock
\showISBNx{9781450394215}
\urldef\tempurl%
\url{https://doi.org/10.1145/3544548.3580910}
\showDOI{\tempurl}


\bibitem[McDowell(1997)]%
        {mcdowell_doing_1997}
\bibfield{author}{\bibinfo{person}{Linda McDowell}.} \bibinfo{year}{1997}\natexlab{}.
\newblock \showarticletitle{‘{Doing} {Gender}: {Feminism}, {Feminists} and {Research} {Methods} in {Human} {Geography}'}.
\newblock In \bibinfo{booktitle}{\emph{Space, {Gender}, {Knowledge}: {Feminist} {Readings}}}. \bibinfo{publisher}{Routledge}.
\newblock
\showISBNx{978-1-315-82487-1}
\newblock
\shownote{Num Pages: 10}.


\bibitem[Mejias and Couldry(2019)]%
        {mejias2019datafication}
\bibfield{author}{\bibinfo{person}{Ulises~A Mejias} {and} \bibinfo{person}{Nick Couldry}.} \bibinfo{year}{2019}\natexlab{}.
\newblock \showarticletitle{Datafication}.
\newblock \bibinfo{journal}{\emph{Internet policy review}} \bibinfo{volume}{8}, \bibinfo{number}{4} (\bibinfo{year}{2019}), \bibinfo{pages}{1--10}.
\newblock


\bibitem[Menking and Rosenberg(2021)]%
        {menking2021wp}
\bibfield{author}{\bibinfo{person}{Amanda Menking} {and} \bibinfo{person}{Jon Rosenberg}.} \bibinfo{year}{2021}\natexlab{}.
\newblock \showarticletitle{WP: NOT, WP: NPOV, and other stories Wikipedia tells us: A feminist critique of Wikipedia’s epistemology}.
\newblock \bibinfo{journal}{\emph{Science, Technology, \& Human Values}} \bibinfo{volume}{46}, \bibinfo{number}{3} (\bibinfo{year}{2021}), \bibinfo{pages}{455--479}.
\newblock


\bibitem[Meyer and Dykes(2020)]%
        {meyer-dykes2020}
\bibfield{author}{\bibinfo{person}{Miriah Meyer} {and} \bibinfo{person}{Jason Dykes}.} \bibinfo{year}{2020}\natexlab{}.
\newblock \showarticletitle{Criteria for Rigor in Visualization Design Study}.
\newblock \bibinfo{journal}{\emph{IEEE Transactions on Visualization and Computer Graphics}} \bibinfo{volume}{26}, \bibinfo{number}{1} (\bibinfo{year}{2020}), \bibinfo{pages}{87--97}.
\newblock
\urldef\tempurl%
\url{https://doi.org/10.1109/TVCG.2019.2934539}
\showDOI{\tempurl}


\bibitem[Morais et~al\mbox{.}(2020)]%
        {Morais2020}
\bibfield{author}{\bibinfo{person}{Luiz Morais}, \bibinfo{person}{Yvonne Jansen}, \bibinfo{person}{Nazareno Andrade}, {and} \bibinfo{person}{Pierre Dragicevic}.} \bibinfo{year}{2020}\natexlab{}.
\newblock \showarticletitle{{Showing Data about People: A Design Space of Anthropographics}}.
\newblock \bibinfo{journal}{\emph{IEEE Transactions on Visualization and Computer Graphics}} \bibinfo{volume}{28}, \bibinfo{number}{3} (\bibinfo{year}{2020}), \bibinfo{pages}{1661--1679}.
\newblock
\urldef\tempurl%
\url{https://doi.org/10.1109/TVCG.2020.3023013}
\showDOI{\tempurl}


\bibitem[Morais et~al\mbox{.}(2021)]%
        {Morais2021}
\bibfield{author}{\bibinfo{person}{Luiz Morais}, \bibinfo{person}{Yvonne Jansen}, \bibinfo{person}{Nazareno Andrade}, {and} \bibinfo{person}{Pierre Dragicevic}.} \bibinfo{year}{2021}\natexlab{}.
\newblock \showarticletitle{{Can Anthropographics Promote Prosociality? A Review and Large-Sample Study}}. In \bibinfo{booktitle}{\emph{Proceedings of the 2021 CHI Conference on Human Factors in Computing Systems}} (Yokohama, Japan) \emph{(\bibinfo{series}{CHI '21})}. \bibinfo{publisher}{Association for Computing Machinery}, \bibinfo{address}{New York, NY, USA}, Article \bibinfo{articleno}{611}, \bibinfo{numpages}{18}~pages.
\newblock
\showISBNx{9781450380966}
\urldef\tempurl%
\url{https://doi.org/10.1145/3411764.3445637}
\showDOI{\tempurl}


\bibitem[Munzner(2009)]%
        {MunznerNested2009}
\bibfield{author}{\bibinfo{person}{Tamara Munzner}.} \bibinfo{year}{2009}\natexlab{}.
\newblock \showarticletitle{A Nested Model for Visualization Design and Validation}.
\newblock \bibinfo{journal}{\emph{IEEE Transactions on Visualization and Computer Graphics}} \bibinfo{volume}{15}, \bibinfo{number}{6} (\bibinfo{year}{2009}), \bibinfo{pages}{921--928}.
\newblock
\urldef\tempurl%
\url{https://doi.org/10.1109/TVCG.2009.111}
\showDOI{\tempurl}


\bibitem[Muth(2024)]%
        {Datawrapper-Race}
\bibfield{author}{\bibinfo{person}{Lisa~Charlotte Muth}.} \bibinfo{year}{2024}\natexlab{}.
\newblock \bibinfo{title}{{What to Consider When Choosing Colors for Race, Ethnicity, and World Regions}}.
\newblock \bibinfo{howpublished}{Available: \url{https://blog.datawrapper.de/colors-for-race-ethnicity-world-regions/}}.
\newblock


\bibitem[{NPR}(2024)]%
        {MENA2024}
\bibfield{author}{\bibinfo{person}{{NPR}}.} \bibinfo{year}{2024}\natexlab{}.
\newblock \bibinfo{title}{{Next U.S. census will have new boxes for 'Middle Eastern or North African,' 'Latino'}}.
\newblock
\newblock
\urldef\tempurl%
\url{https://www.npr.org/2024/03/28/1237218459/census-race-categories-ethnicity-middle-east-north-africa}
\showURL{%
Retrieved September 12, 2024 from \tempurl}


\bibitem[Ogbonnaya-Ogburu et~al\mbox{.}(2020)]%
        {Ogbonnaya2020}
\bibfield{author}{\bibinfo{person}{Ihudiya~Finda Ogbonnaya-Ogburu}, \bibinfo{person}{Angela~D.R. Smith}, \bibinfo{person}{Alexandra To}, {and} \bibinfo{person}{Kentaro Toyama}.} \bibinfo{year}{2020}\natexlab{}.
\newblock \showarticletitle{{Critical Race Theory for HCI}}. In \bibinfo{booktitle}{\emph{Proceedings of the 2020 CHI Conference on Human Factors in Computing Systems}} (Honolulu, HI, USA) \emph{(\bibinfo{series}{CHI '20})}. \bibinfo{publisher}{Association for Computing Machinery}, \bibinfo{address}{New York, NY, USA}, \bibinfo{pages}{1–16}.
\newblock
\showISBNx{9781450367080}
\urldef\tempurl%
\url{https://doi.org/10.1145/3313831.3376392}
\showDOI{\tempurl}


\bibitem[O’Connor et~al\mbox{.}(2023)]%
        {oconnor2023decolonizing}
\bibfield{author}{\bibinfo{person}{January O’Connor}, \bibinfo{person}{Mark Parman}, \bibinfo{person}{Nicole Bowman}, {and} \bibinfo{person}{Stephanie Evergreen}.} \bibinfo{year}{2023}\natexlab{}.
\newblock \showarticletitle{Decolonizing data visualization: A history and future of Indigenous data visualization}.
\newblock \bibinfo{journal}{\emph{Journal of MultiDisciplinary Evaluation}} \bibinfo{volume}{19}, \bibinfo{number}{44} (\bibinfo{year}{2023}), \bibinfo{pages}{62--79}.
\newblock


\bibitem[Parsons(2021)]%
        {parsons2021understanding}
\bibfield{author}{\bibinfo{person}{Paul Parsons}.} \bibinfo{year}{2021}\natexlab{}.
\newblock \showarticletitle{Understanding data visualization design practice}.
\newblock \bibinfo{journal}{\emph{IEEE Transactions on Visualization and Computer Graphics}} \bibinfo{volume}{28}, \bibinfo{number}{1} (\bibinfo{year}{2021}), \bibinfo{pages}{665--675}.
\newblock
\urldef\tempurl%
\url{https://doi.org/10.1109/TVCG.2021.3114959}
\showDOI{\tempurl}


\bibitem[Parsons et~al\mbox{.}(2021)]%
        {parsons2021fixation}
\bibfield{author}{\bibinfo{person}{Paul Parsons}, \bibinfo{person}{Prakash Shukla}, {and} \bibinfo{person}{Chorong Park}.} \bibinfo{year}{2021}\natexlab{}.
\newblock \showarticletitle{Fixation and Creativity in Data Visualization Design: Experiences and Perspectives of Practitioners}. In \bibinfo{booktitle}{\emph{2021 IEEE Visualization Conference (VIS)}}. IEEE Computer Society, \bibinfo{publisher}{IEEE}, \bibinfo{pages}{76--80}.
\newblock


\bibitem[Peck et~al\mbox{.}(2019)]%
        {peck}
\bibfield{author}{\bibinfo{person}{Evan~M. Peck}, \bibinfo{person}{Sofia~E. Ayuso}, {and} \bibinfo{person}{Omar El-Etr}.} \bibinfo{year}{2019}\natexlab{}.
\newblock \showarticletitle{{Data is Personal: Attitudes and Perceptions of Data Visualization in Rural Pennsylvania}}. In \bibinfo{booktitle}{\emph{Proceedings of the 2019 CHI Conference on Human Factors in Computing Systems}} (Glasgow, Scotland UK) \emph{(\bibinfo{series}{CHI '19})}. \bibinfo{publisher}{Association for Computing Machinery}, \bibinfo{address}{New York, NY, USA}, \bibinfo{pages}{1–12}.
\newblock
\showISBNx{9781450359702}
\urldef\tempurl%
\url{https://doi.org/10.1145/3290605.3300474}
\showDOI{\tempurl}


\bibitem[Pine and Liboiron(2015)]%
        {pine2015politics}
\bibfield{author}{\bibinfo{person}{Kathleen~H. Pine} {and} \bibinfo{person}{Max Liboiron}.} \bibinfo{year}{2015}\natexlab{}.
\newblock \showarticletitle{The Politics of Measurement and Action}. In \bibinfo{booktitle}{\emph{Proceedings of the 33rd Annual ACM Conference on Human Factors in Computing Systems}} (Seoul, Republic of Korea) \emph{(\bibinfo{series}{CHI '15})}. \bibinfo{publisher}{Association for Computing Machinery}, \bibinfo{address}{New York, NY, USA}, \bibinfo{pages}{3147–3156}.
\newblock
\showISBNx{9781450331456}
\urldef\tempurl%
\url{https://doi.org/10.1145/2702123.2702298}
\showDOI{\tempurl}


\bibitem[Renzaho(2023)]%
        {renzaho2023lack}
\bibfield{author}{\bibinfo{person}{Andre~MN Renzaho}.} \bibinfo{year}{2023}\natexlab{}.
\newblock \showarticletitle{The lack of race and ethnicity data in Australia—a threat to achieving health equity}.
\newblock \bibinfo{journal}{\emph{International Journal of Environmental Research and Public Health}} \bibinfo{volume}{20}, \bibinfo{number}{8} (\bibinfo{year}{2023}), \bibinfo{pages}{5530}.
\newblock


\bibitem[Scheuerman et~al\mbox{.}(2021)]%
        {scheuerman2021datasets}
\bibfield{author}{\bibinfo{person}{Morgan~Klaus Scheuerman}, \bibinfo{person}{Alex Hanna}, {and} \bibinfo{person}{Emily Denton}.} \bibinfo{year}{2021}\natexlab{}.
\newblock \showarticletitle{Do Datasets Have Politics? Disciplinary Values in Computer Vision Dataset Development}.
\newblock \bibinfo{journal}{\emph{Proceedings of the ACM on Human-Computer Interaction}} \bibinfo{volume}{5}, \bibinfo{number}{CSCW2} (\bibinfo{year}{2021}), \bibinfo{pages}{1--37}.
\newblock


\bibitem[Schwabish and Feng(2021)]%
        {SchwabishFeng}
\bibfield{author}{\bibinfo{person}{Jonathan Schwabish} {and} \bibinfo{person}{Alice Feng}.} \bibinfo{year}{2021}\natexlab{}.
\newblock \bibinfo{title}{{Do No Harm Guide: Applying Equity Awareness in Data Visualization}}.
\newblock
\newblock
\newblock
\shownote{Urban Institute. \url{https://www. urban. org/research/publication/do-no-harm-guide-applying-equity-awareness-data-visualization}}.


\bibitem[Schwan et~al\mbox{.}(2022)]%
        {Schwan2022}
\bibfield{author}{\bibinfo{person}{Hannah Schwan}, \bibinfo{person}{Jonas Arndt}, {and} \bibinfo{person}{Marian Dörk}.} \bibinfo{year}{2022}\natexlab{}.
\newblock \showarticletitle{{Disclosure as a critical-feminist design practice for Web-based data stories}}.
\newblock \bibinfo{journal}{\emph{First Monday}} \bibinfo{volume}{27}, \bibinfo{number}{11} (\bibinfo{date}{Nov.} \bibinfo{year}{2022}), \bibinfo{numpages}{27}~pages.
\newblock
\urldef\tempurl%
\url{https://doi.org/10.5210/fm.v27i11.12712}
\showDOI{\tempurl}


\bibitem[Spiel et~al\mbox{.}(2019)]%
        {Spiel2019}
\bibfield{author}{\bibinfo{person}{Katta Spiel}, \bibinfo{person}{Oliver~L. Haimson}, {and} \bibinfo{person}{Danielle Lottridge}.} \bibinfo{year}{2019}\natexlab{}.
\newblock \showarticletitle{How to do better with gender on surveys: a guide for HCI researchers}.
\newblock \bibinfo{journal}{\emph{Interactions}} \bibinfo{volume}{26}, \bibinfo{number}{4} (\bibinfo{date}{June} \bibinfo{year}{2019}), \bibinfo{pages}{62–65}.
\newblock
\showISSN{1072-5520}
\urldef\tempurl%
\url{https://doi.org/10.1145/3338283}
\showDOI{\tempurl}


\bibitem[Tovanich et~al\mbox{.}(2022)]%
        {tovanich2022}
\bibfield{author}{\bibinfo{person}{Natkamon Tovanich}, \bibinfo{person}{Pierre Dragicevic}, {and} \bibinfo{person}{Petra Isenberg}.} \bibinfo{year}{2022}\natexlab{}.
\newblock \showarticletitle{Gender in 30 Years of IEEE Visualization}.
\newblock \bibinfo{journal}{\emph{IEEE Transactions on Visualization and Computer Graphics}} \bibinfo{volume}{28}, \bibinfo{number}{1} (\bibinfo{year}{2022}), \bibinfo{pages}{497--507}.
\newblock
\urldef\tempurl%
\url{https://doi.org/10.1109/TVCG.2021.3114787}
\showDOI{\tempurl}


\bibitem[{United States Equal Employment Opportunity Commission}(2023)]%
        {ProtectedDem}
\bibfield{author}{\bibinfo{person}{{United States Equal Employment Opportunity Commission}}.} \bibinfo{year}{2023}\natexlab{}.
\newblock \bibinfo{title}{{Who is protected from employment discrimination?}}
\newblock
\newblock
\urldef\tempurl%
\url{https://perma.cc/5XAQ-9763}
\showURL{%
Retrieved September 12, 2024 from \tempurl}


\bibitem[Winner(1980)]%
        {Winner1980-WINDAH-3}
\bibfield{author}{\bibinfo{person}{Langdon Winner}.} \bibinfo{year}{1980}\natexlab{}.
\newblock \showarticletitle{Do Artifacts Have Politics?}
\newblock \bibinfo{journal}{\emph{Daedalus}} \bibinfo{volume}{109}, \bibinfo{number}{1} (\bibinfo{year}{1980}), \bibinfo{pages}{121--136}.
\newblock


\bibitem[Zhang et~al\mbox{.}(2022)]%
        {zhang2022visualization}
\bibfield{author}{\bibinfo{person}{Yixuan Zhang}, \bibinfo{person}{Yifan Sun}, \bibinfo{person}{Joseph~D Gaggiano}, \bibinfo{person}{Neha Kumar}, \bibinfo{person}{Clio Andris}, {and} \bibinfo{person}{Andrea~G Parker}.} \bibinfo{year}{2022}\natexlab{}.
\newblock \showarticletitle{Visualization design practices in a crisis: Behind the scenes with COVID-19 dashboard creators}.
\newblock \bibinfo{journal}{\emph{IEEE Transactions on Visualization and Computer Graphics}} \bibinfo{volume}{29}, \bibinfo{number}{1} (\bibinfo{year}{2022}), \bibinfo{pages}{1037--1047}.
\newblock


\bibitem[Zraick et~al\mbox{.}(2024)]%
        {MENA-NYT}
\bibfield{author}{\bibinfo{person}{Karen Zraick}, \bibinfo{person}{Allison McCann}, \bibinfo{person}{Sarah Almukhtar}, \bibinfo{person}{Yuliya Parshina-kottas}, \bibinfo{person}{Robert Gebeloff}, {and} \bibinfo{person}{Denise Lu}.} \bibinfo{year}{2024}\natexlab{}.
\newblock \bibinfo{title}{{No Box to Check: When the Census Doesn't Reflect You}}.
\newblock \bibinfo{howpublished}{Available: \url{https://www.nytimes.com/interactive/2024/02/25/us/census-race-ethnicity-middle-east-north-africa.html}}.
\newblock


\end{thebibliography}

% \section{Appendix}
% https://tex.stackexchange.com/a/259296
%TC:ignore 
\appendix
\section{Semi-Structured Interview Protocol}\label{appendix:protocol}

\textbf{We’ll start with a few questions about your background.} 
\begin{itemize}
    \item Tell me about your professional path to your current role. How long have you been working as a visualization designer? 
    \item Tell me about the kinds of data visualization work you do in your current position or your most recent position. What does your day to day look like? 
\end{itemize}

\textbf{Now, I would like us to talk about your design experiences.}
\begin{itemize}
    \item  Think back to an experience designing data visualizations that have included some aspects of race or gender demographic categories. If it is disclosable, can you share the visualization or a rough sketch with me? Do I have your consent to include this visualization in future publications of this study? 
    \item How did you decide how to represent the data for these categories?
    \item How do you handle missing data or other data issues related to protected demographic categories in your design process?
    \item How do you think these issues in the underlying data shape your design process?
\end{itemize} 

\textbf{I saw that you mentioned that you used [participant response] tools in the recruitment questionnaire.}
\begin{itemize}
    \item Do these tools meet your design needs for data about race and other demographic categories that you have worked with? 
    \item What challenges, if any, have you experienced when trying to use these tools? 
\end{itemize}
\textbf{Let’s move on to your process and design values.}
\begin{itemize}
    \item Can you describe your design process for creating these kinds of data visualizations? 
    \item What design values or best practices do you use in your design process? How? 
    \item How does your design process change (if it does) depending on the intended audiences of the data visualizations? How much power do you believe you have in making these decisions?
    \item What are the primary design values that come up when you are designing data visualizations? What does that value mean to you? 
    \item How would you describe the identities that are most important to you and your politics?
How do your personal beliefs, values, biases or politics influence your design process? [Let the participant respond] Can you give me some examples?
\end{itemize}

\textbf{Please feel free to not answer this next question.}
\begin{itemize}
    \item Have you ever created a data visualization about race and other protected categories on an issue that you disagreed with or had differing opinions from your organization?
    \item How do you think that experience differed from your usual design process? 
    \item How do you understand neutrality in your design process in such a situation? 
    \item Can you share a visualization artifact that you designed? How would you redesign this visualization if you did not have any restrictions?
    \item Are there any resources that you would recommend to someone getting started in visualizing these kinds of protected demographic categories?
\end{itemize}

\textbf{Wrapping up the interview}

We’re now nearing the end of the interview. As we wrap up, is there anything else that you would like to share with me on your experiences designing data visualizations of race and other protected demographic categories? 
%TC:endignore 

\end{document}